\documentclass[traditabstract]{aa} 
\usepackage{graphicx}

\usepackage{txfonts}

\usepackage{caption}
\usepackage{natbib}
\newcommand{\bd}{\begin{displaymath}}
\newcommand{\ed}{\end{displaymath}}
\newcommand{\be}{\begin{equation}}
\newcommand{\ee}{\end{equation}}
\newcommand{\beaa}{\begin{eqnarray*}}
\newcommand{\eeaa}{\end{eqnarray*}}
\newcommand{\bea}{\begin{eqnarray}}
\newcommand{\eea}{\end{eqnarray}}

\def\ourlens{MACS\,J0416.1$-$2403{}}

\def\zc{z_{\rm c}}

\def\macs{MACS0416}

\begin{document}
   \title{\ourlens : Impact of line-of-sight structures on strong gravitational lensing modelling of galaxy clusters}

  \titlerunning{\ourlens : Impact of line-of-sight structures on clusters strong gravitational lensing modelling}

   \author{G. Chiriv\`i\inst{1}
   		 \and    
          S. H. Suyu\inst{1,2,3}
                    \and
          C. Grillo \inst{4,5}  
                  \and 
          A. Halkola        
               \and
                I. Balestra\inst{6,7}
             \and 
             G. B. Caminha\inst{8,9}
             \and             
           A. Mercurio\inst{10}
             \and 
             P. Rosati\inst{8}
          }

   \institute{Max-Planck-Institut f\"ur Astrophysik, Karl-Schwarzschild Str. 1, 85741 Garching, Germany\\
              \email{chirivig@MPA-Garching.MPG.DE}
         \and 
           Institute of Astronomy and Astrophysics, Academia Sinica, P.O.~Box 23-141, Taipei 10617, Taiwan  
           \and 
           Physik-Department, Technische Universit\"at M\"unchen, James-Franck-Stra\ss{}e~1, 85748 Garching, Germany
           \and
           Dipartimento di Fisica, Universit\`a degli Studi di Milano, via Celoria 16, I-20133 Milano, Italy
           \and
           Dark Cosmology Centre, Niels Bohr Institute, University of Copenhagen, Juliane Maries Vej 30, DK-2100 Copenhagen, Denmark    
           \and
           University Observatory Munich, Scheinerstrasse 1, 81679 Munich, Germany
           \and
         INAF - Osservatorio Astronomico di Trieste, via G. B. Tiepolo 11, I-34143, Trieste, Italy
            \and
           Dipartimento di Fisica e Scienze della Terra, Universit\`a degli Studi di Ferrara, Via Saragat 1, I-44122 Ferrara, Italy
           \and
           Osservatorio di Bologna INAF - Osservatorio Astronomico di Bologna, Via Ranzani 1, I- 40127 Bologna, Italy
			\and
           INAF - Osservatorio Astronomico di Capodimonte, Via Moiariello 16, I-80131 Napoli, Italy\\
             }

   \date{Received --; accepted --}

 
  \abstract
  {Exploiting the powerful tool of strong gravitational lensing by galaxy clusters to study the highest-redshift Universe and cluster mass distributions relies on precise lens mass modelling. In this work, we present the first attempt at modelling line-of-sight mass distribution in addition to that of the cluster, extending previous modelling techniques that assume mass distributions to be on a single lens plane. We focus on the Hubble Frontier Field cluster \ourlens, and our multi-plane model reproduces the observed image positions with a rms offset of $\sim0\farcs53$. Starting from this best-fitting model, we simulate a mock cluster that resembles \ourlens\ in order to explore the effects of line-of-sight structures on cluster mass modelling.  By systematically analysing the mock cluster under different model assumptions, we find that neglecting the lensing environment has a significant impact on the reconstruction of image positions (rms $\sim 0\farcs3$); accounting for line-of-sight galaxies as if they were at the cluster redshift can partially reduce this offset. Moreover, foreground galaxies are more important to include into the model than the background ones. While the magnification factor of the lensed multiple images are recovered within $\sim10\%$ for $\sim95\%$ of them, those $\sim 5\%$ that lie near critical curves can be significantly affected by the exclusion of the lensing environment in the models (up to a factor of ~200). In addition, line-of-sight galaxies cannot explain the apparent discrepancy in the properties of massive subhalos between \ourlens\ and N-body simulated clusters. Since our model of \ourlens\ with line-of-sight galaxies only reduced modestly the rms offset in the image positions, we conclude that additional complexities, such as more flexible halo shapes, would be needed in future models of \ourlens.
}
   {}
   {}
   {}
   {}


   \maketitle
%

\section{Introduction}
\label{sec:intro}

Massive galaxy clusters are the largest gravitationally bound structures in the Universe and they are located at the nodes of the Cosmic Web. According to the currently accepted cosmological model, which consists of a cold dark matter dominated Universe with a cosmological constant ($\rm \Lambda CDM$), more massive structures form by accretion and assembly of smaller self-bound individual systems \citep[e.g.][]{Springel2006}.  As such, galaxy clusters are not only a perfect laboratory to study the formation and evolution of structures in the Universe \citep[e.g.][and references therein]{Dressler1984, KravtsovBorgani2012}, but also to study the mass-energy density components, as dark matter and dark energy, and to constrain cosmological parameters \citep[e.g.][among others]{Jullo2010, Caminha2016, Rozo2010, Planck2014}. \\
They are also very efficient cosmic telescopes. Indeed, the magnification effect produced by gravitational lensing with galaxy clusters provides a powerful tool to detect and study high-redshift galaxies, that would be undetectable with currently available instruments. Gravitational lensing is a relativistic effect for which the light travelling from a source towards the observer is bent by the presence of matter in-between. Consequently, the source will be observed at a different position than it actually is, distorted in shape, and in some cases also multiply imaged (in the so-called $``$strong lensing$"$ regime). It will also appear magnified by a factor $\mu$ (magnification). Once the magnification $\mu$ is known, due to surface brightness conservation, the intrinsic brightness and shape of the source galaxy can be reconstructed. This will allow one to study high-redshift galaxies, providing crucial probes of structure formation and galaxy evolution. \\
In the past decades, gravitational lensing with clusters has highly improved our knowledge of the mass distribution in clusters, and has led to the discovery of some of the highest-redshift galaxies \citep[e.g.][]{Coe2013, Bouwens2014}.
Being such versatile laboratories for many studies, galaxy clusters have been searched for and studied in depth in recent years. In particular, the Hubble Frontier Fields initiative (HFF; P.I.: J. Lotz) has exploited the $\it Hubble\  Space\ Telescope\ (HST)$ sensitivity and the magnification effect of six targeted strong lensing clusters to detect the high-redshift Universe. These six clusters were observed in seven optical and near-IR bands using the Advanced Camera for Survey\ (ACS) and the Wide Field Camera 3 (WFC3), achieving an unprecedent depth of $\sim 29$ mag (AB). The HFF initiative has provided a sample of high-redshift galaxies that will allow one to investigate their properties in a statistically significant way. \\
As previously mentioned, to reconstruct the intrinsic brightness of these far away sources, the magnification effect needs to be accounted for. This can be done only by modelling in detail the mass distribution of the lensing cluster. High precision reconstruction of cluster lenses is necessary to avoid systematic errors affecting the cluster mass reconstruction or the source brightness substantially.
In recent years, the models of clusters of galaxies have been performed with increasing precision \citep[e.g.][among others]{Grillo2015,Caminha2016}. However, current analyses using single-plane approach seem to have reached a limit in reproducing the observables. This approach consists of modelling the lens as if it were an isolated system on a single redshift-plane, and has led to models with an rms between the observed and predicted positions of the multiple images of the strongly lensed background sources of $\sim 0\farcs5-1''$, which is greater than the observational uncertainty  ($\sim 0\farcs05$). To account for the residual offset, it might be necessary to also consider the lensing environment in the model. In fact, we expect that the contribution of the line-of-sight (LOS) matter can affect the observables more than the observational uncertainties. Therefore, this effect needs to be taken into account to study its contribution to the recovered offset between observations and current cluster models \citep[e.g.][]{DaloisioNatarajan2014, CaminhaGrillo2016}.\\
In this work we simulate galaxy cluster lensing observations, accounting for the LOS effects using multi-plane lensing formalism, and study the effects of the LOS structures on the magnification and position of the images. We then analyse these simulated data to study the cluster mass distribution and to reconstruct the high-redshift galaxies’ intrinsic brightness. We analyse toy models of clusters, and then a mock model that is as similar as possible to the HFF cluster \ourlens , to make our study more realistic.\\
The work is organised as follows: in Section \ref{sec:MPlm} we describe our analysis' method, the multi-plane formalism, and the profiles we use to parametrise the lens cluster. In Section \ref{sec:toymod} we analyse toy models of simplistic clusters and in Section \ref{sec:0416} we model the mass distribution of a real cluster, namely the HFF cluster \ourlens . We use our best-fit model to generate lensing observables from a simulated mock system that mimics \ourlens\ and we model the simulated cluster with different assumptions. We discuss our results in Section \ref{sec:summary}.\\
Throughout this work, we assume a flat $\Lambda$-CDM cosmology with $H_0=70 \rm{\, km\, s^{-1}\, Mpc^{-1}}$ and $\Omega_{\Lambda}=1-\Omega_{\rm M}=0.70$.  From the redshift of the lens ($z_{\rm c}=0.396$) in Section \ref{sec:MACSintro}, one arcsecond at the lens plane in \ourlens\ corresponds to $\sim5.34\ {\rm \, kpc}$.  The magnitudes are all given in the AB system. The uncertainty we consider for all the image positions corresponds to the pixel scale $0\farcs065$, unless stated otherwise.


\section{Multiple lens plane modelling}
\label{sec:MPlm}
In this work, the modelling is obtained using GLEE, a software developed by A. Halkola and S. H. Suyu  \citep{SuyuHalkola2010, Suyu2012}. This software uses parametrised mass profiles to describe the halo and galaxies and a Bayesian analysis to infer the best-fit parameter values and their variances and degeneracies. It also includes the possibility of considering lenses at different redshifts (multiple lens plane modelling; Suyu et al. in preparation). In the following sections we introduce the multiple lens plane formalism (Section \ref{sec:MP formalism}), we describe the lens profiles that we use for the lens galaxies and dark matter halos (Section \ref{Lensprofiles}) and the scaling relations we use for the cluster members and the LOS galaxies (Section \ref{sec:scalrel}). We discuss how we determine the model parameters using a Bayesian approach in Section \ref{sec:modelpar}.

\subsection{Multi-plane Formalism}
\label{sec:MP formalism}
In this Section we briefly revisit the generalised multi-plane gravitational lens theory  \citep[e.g.][]{Blandford&Natarajan1986, Schneider1992}, which considers the fact that a light ray can be bent multiple times by several deflectors during its path. This theory takes into account the effect of secondary lenses at different redshifts using the thin-lens approximation for every deflector on its redshift plane. The lens equation in this formalism is \citep[following][]{Gavazzi2008}

\be
\label{eq:MPlenseqn}
$$\vec{\theta}_j (\vec{\theta}_{\rm 1})= \vec{\theta}_{\rm 1}- \sum_{i=1}^{j-1} \frac{D_{ij}}{D_j} \vec{\alpha}(\vec{\theta}_i), $$        
\ee
where $\vec{\theta}_{\rm 1}$ is the image position on the 1st plane (observed image plane), $\vec{\theta}_j$ is the image position on the $j$-th plane and $\vec{\alpha}(\vec{\theta}_i)$ is the deflection angle on the $i$-th plane. In this recursive equation the deflection angle on one plane depends on the deflection angle of all the previous planes. The source position $\vec{\beta}$, which is on the $N$-th plane, corresponds to

\be
\label{eq:MPbeta}
$$\vec{\beta}= \vec{\theta}_{N} (\vec{\theta}_{\rm 1})= \vec{\theta}_{\rm 1}- \sum_{i=1}^{N-1} \frac{D_{i  N}}{D_{ N}} \vec{\alpha}(\vec{\theta}_i).         $$ 
\ee
Therefore the total deflection angle $\vec{\alpha}_{\rm tot}$ is the sum of all the deflection angles on all planes, namely 

\be
\label{eq:MPalphatot}
$$\vec{\alpha}_{\rm tot}= \sum_{i=1}^{N-1} \frac{D_{i N}}{D_{ N}} \vec{\alpha}(\vec{\theta}_i).         $$
\ee

\subsection{Lens mass distributions}
\label{Lensprofiles}
In GLEE, we use parametric mass profiles to portray the cluster members component, the line-of-sight-perturber galaxies component, and the contribution of the remaining intra cluster mass (mainly dark matter) respectively.
We model the luminous mass component (i.e. members and LOS galaxies) with a truncated dual pseudo-isothermal elliptical mass distribution \citep[dPIE;][]{EliasdottirEtal07, SuyuHalkola2010} with vanishing core radius. Their dimensionless projected surface mass density, i.e. convergence, for a source at $z_{\rm s}=\infty$, is:

\begin{equation}
\label{kappadPIE}
\scalebox{1.3}{$ \kappa_{\rm dpie}(x,y)= \frac{\theta_{\rm E}}{2} \left( \frac{1}{R_{\rm em}} - \frac{1}{\sqrt{R_{\rm em}^2+r_{\rm t}^2}}\right) ,$}
\end{equation}
where $(x, y)$ are the coordinates on the lens plane, $\theta_{\rm E}$ is the lens Einstein radius, $r_{\rm t}$ is the truncation radius. The mass distribution is then suitably rotated by its orientation angle $\theta$ and shifted by the centroid position of the coordinate system used. The 2D elliptical mass radius is

\begin{equation}
\label{rem}
\scalebox{1.3}{$ R_{\rm em}=\sqrt{\frac{x^2}{(1+e)^2}+\frac{y^2}{(1-e)^2}}, $}
\end{equation}
and the ellipticity is

\begin{equation}
\label{e}
\scalebox{1.3}{$e=\frac{1-q}{1+q}$}
 \end{equation}
 where $q$ is the axis ratio. Therefore, the parameters that identify this profile are its central position $(x_{\rm c}, y_{\rm c})$, its axis ratio $q$, its orientation $\theta$, its strength $\theta_{\rm E}$, and its truncation radius $r_{\rm t}$.
The 3D density corresponding to this $\kappa$ is 

\begin{equation}
\label{rhodPIE}
\scalebox{1.1}{$ \rho_{\rm dpie}(r)\propto \left(r^2(r^2+r_{\rm t}^2)\right)^{-1}, $}
\end{equation}
where $r$ is the 3D radius. As we can see, for $r$ greater than the truncation radius, this density distribution is truncated, i.e. scales as $r^{-4}$. To be noted in isothermal profiles there is a direct relation between the value of the velocity dispersion $\sigma$ of the lens and that of its strength $\theta_{\rm E}$, namely

\begin{equation}
\label{SISveldisp}
\scalebox{1.3}{$ \frac{\sigma}{c}= \sqrt{\frac{\theta_{\rm E}}{4 \pi}}.$}
\end{equation}
To model the remaining mass of the cluster, especially the contribute of dark matter halos, we use a 2D pseudo-isothermal elliptical mass distribution \citep[PIEMD; ][]{KassiolaKovner93} or a softened power-law elliptical mass distributions \citep[SPEMD; ][]{Barkana98}. The convergence of the PIEMD profile is 

\begin{equation}
\label{kappaPIEMD}
\scalebox{1.3}{$ \kappa_{\rm piemd}(x,y)= \frac{\theta_{\rm E}}{2 \sqrt{R_{\rm em}^2 +r_{\rm c}^2}},  $}
\end{equation}
where, again, $(x,y)$ are the coordinates in the lens plane, $ R_{\rm em}$ is the elliptical mass radius,  $\theta_{\rm E}$ is the lens Einstein radius and $r_{\rm c}$ is the core radius.  The mass distribution is then suitably rotated by its orientation angle $\theta$ and shifted by the centroid position of the coordinate system used. The parameters that identify this profile are its central position $(x_{\rm c}, y_{\rm c})$, its axis ratio $q$, its orientation $\theta$, its strenght $\theta_{\rm E}$, and its core radius $r_{\rm c}$.
In the case of the SPEMD, there is one additional parameter, which is the slope $\gamma$ of the profile, since its convergence is 

\begin{equation}
\label{kappaSPEMD}
\scalebox{1.3}{$ \kappa_{\rm spemd}(x,y)= \theta_{\rm E} \left( x^2+\frac{y^2}{q^2}+r_{\rm c}^2 \right) ^{-\gamma},  $}
\end{equation}
where $q$ is the axis ratio, $r_{\rm c}$ is the core radius, $\gamma$ is the power law index, which is 0.5 for an isothermal profile  \citep{Barkana98}.

\subsection{Scaling relation}
\label{sec:scalrel}
Throughout this work, to reduce the number of model parameters and therefore increase the computational efficiency of our modelling, we assume scaling relations for the Einstein radii and truncation radii of the cluster members, i.e. we scale them with respect to a reference galaxy, which we choose as the brightest cluster galaxy (BCG). The scaling is done following \citet{Grillo2015},
\be
\label{eq:ML_tiltscale}
\scalebox{1.1}{ $\theta_{\rm E, \it i} \propto \theta_{\rm E, g} \left(\frac{L_i}{L_{\rm g}}\right)^{0.7}, \\ r_{\rm t, \it i} \propto r_{\rm t,g} \left(\frac{L_i}{L_{\rm g}}\right)^{0.5},$}
\ee
where $\theta_{\rm E,\it i}$ and $r_{\rm t,\it i}$ are the Einstein radius and the truncation radius of the $i$-th galaxy with luminosity $L_i$, and $\theta_{\rm E,g}$, $r_{\rm t,g}$, $L_{\rm g}$ are the properties of the BCG. This is equivalent to having cluster members with a total mass-to-light ratio $\frac{M_{\rm T}}{L} \propto L^{0.2}$ , also known as the tilt of the fundamental plane \citep{Faber1987,Bender1992}, since

\be
\label{eq:ML_tiltscaleproof}
$$\frac{M_{\rm T, \it i}}{L_i} \sim \frac{\sigma_i^2 r_{\rm t, \it i}}{L_i} \sim \frac{L_i^{0.7}L_i^{0.5}}{L_i} \sim L_i^{0.2}.$$
\ee
To test how the values of the parameters change if we assume a different scaling relation, we also use
\be
\label{eq:ML_constscale}
\scalebox{1.1}{$\theta_{\rm E,\it i} \propto \theta_{\rm E,g} \left(\frac{L_i}{L_{\rm g}}\right)^{0.5}, \\ r_{\rm t,\it i} \propto r_{\rm t,g} \left(\frac{L_i}{L_{\rm g}}\right)^{0.5}.$}
\ee
This is equivalent to having cluster members with constant total mass-to-light ratio, since, as shown in \citet{Grillo2015}

\be
\label{eq:ML_const}
$$\frac{M_{\rm T, \it i}}{L_i} \sim \frac{\sigma_i^2 r_{\rm t, \it i}}{L_i} \sim \frac{L_i^{0.5}L_i^{0.5}}{L_i} \sim L_i^{0}.$$
\ee

\subsection{Determination of model parameters}
\label{sec:modelpar}

 GLEE uses a Bayesian analysis to infer the best-fit parameters,  i.e. those that maximise the likelihood on the image plane (given uniform priors on the parameters), namely the probability of the data (the observed image positions $\vec{X}_{\rm obs}$, which corresponds to $\vec{\theta}_1$ in Section \ref{sec:MP formalism}) given a set of parameters $\vec{\eta}$ of the model. This likelihood
\be
\label{eq:likelihood}
$$\mathcal{L}(\vec{X}_{\rm obs}|\vec{\eta})  \propto \exp { \left[ -\frac{1}{2}\sum_{j=1}^{\rm N_{\rm sys}}   \sum_{i=1}^{\rm N_{{\rm im}, j} }      \frac{| \vec{X}_{i,j}^{\rm obs}- \vec{X}_{i,j}^{\rm pred}(\eta) |^2}{\sigma_{i,j}^2}          \right] }$$
\ee
describes the offset between the observed  $\vec{X}_{i,j}^{\rm obs}\ $  and predicted $\vec{X}_{i,j}^{ \rm pred}\ $ image positions for all the images $N_{{\rm im}, j}$ of the  observed image system $j$ of $N_{\rm sys}$, where  $\sigma_{i,j}$ is the observational uncertainty of the $i$-th image of the $j$-th source. We use a simulated annealing technique to find the global minimum and recover the best-fit parameter values, Markov Chain Monte Carlo \citep[MCMC methods based on][]{DunkleyEtal05} and {\sc Emcee} \citep{ForemanMackey16} to sample their posterior probability distributions. The posterior probability is obtained using Bayes' Theorem 

\begin{equation}
\label{bayesthm}
$$ P(\vec{\eta} | \vec{X}_{\rm obs}) \propto \mathcal{L}(\vec{X}_{\rm obs}|\vec{\eta}) P(\vec{\eta}),$$
\end{equation}
where $P(\vec{\eta})$ is the prior probability on the model parameters, that we always assume to be uniform.
The model parameter values we show are the median of the one-dimensional marginalized posterior PDF, with the quoted uncertainties showing the 16th and 84th percentiles (i.e., the bounds of a $68 \%$ credible interval). To obtain realistic uncertainties for the model parameters, we run a second MCMC analysis where we increase the image position uncertainty to roughly the value of the $ rms$, to account for the model imperfections (e.g. lack of treatment for elliptical cluster members, simplistic halo models, neglecting small dark-matter clumps).

\section{Toy Models}
\label{sec:toymod}

We begin with simple toy models of lenses at multiple redshifts in order to gain intuition for multi-plane lens modelling. We generate mock lensing observables and analyse the impact of the introduction of multi-plane lenses by fitting the same image positions with both a multi-plane-lens and a single plane-lens model.

\subsection{Two lenses at different redshifts}
\label{sec:2lenspl}

\begin{figure}
  \centering
   \includegraphics[width=1.0\columnwidth]{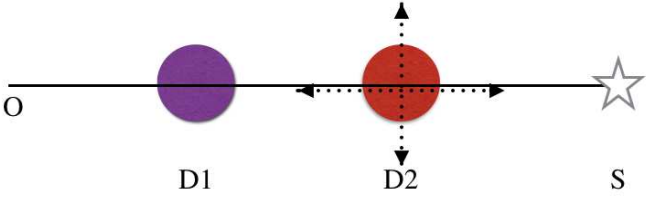} 
   \caption{Positional setup of the two lenses. The first lens (purple) is at $z_{\rm d1}=0.5$, the second lens (red) at $\ z_{\rm d2}= 0.7$. Both the lens are SISs. The dotted arrows indicate the direction in which we shift the second lens to experiment the effects on the image positions and magnification.}
   \label{fig:2lens_setup}
\end{figure}

We start our analysis with a simple toy model composed of two lenses (modelled as SISs), at redshifts $z_{\rm d1}=0.5,\ z_{\rm d2}= 0.7$ aligned along the LOS and with Einstein radii of $\sim2.6 \farcs$ and $\sim1.6\farcs$ respectively, and a point source at redshift $z_{\rm s}=2$, as shown in Figure \ref{fig:2lens_setup}. We study different configurations, in terms of lens location and lensed image positions, and test how well, for the same simulated observed image positions, we are able to reconstruct the properties of the system using a single lens at $z_{\rm sd}=0.5$ whose center is coincident with the input position of SIS1. We find that the variation of the mass of SIS2 still allows us to fit the image and source positions and reconstruct the magnification with a single lens, if the two lenses are aligned. We obtain a $\theta_{\rm E,sd}$ which increases as we increase the mass of the second lens. Instead, variations in the position of SIS2 with respect to the optical axis (we choose a range of $0\farcs2-1\farcs5$) still allow a perfect fit to the same image positions with a single lens, whose $\theta_{\rm E,sd}$ is unchanged with respect to changes of SIS2's distance from the optical axis. However, the magnification of the multi-plane system decreases as SIS2 moves further away from the optical axis. Finally, if we shift SIS2 on the optical axis to higher redshifts, we observe that $\theta_{\rm E,sd}$ decreases, while the magnification increases accordingly for both the multi-plane and single plane lens system.

\subsection{Mock cluster lensing mass distribution 2}
\label{sec:Mock2}

\subsubsection{Input} 
\label{sec:Mock2input}
\begin{figure}[htbp]
   \centering
   \includegraphics[width=1.0\columnwidth]{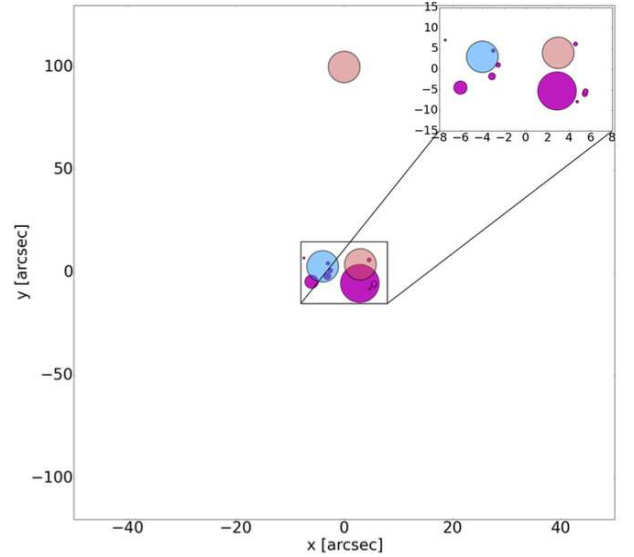} 
   \caption{Cluster galaxies' positional distribution for mock cluster lensing mass distribution 2.  The transparent circles are the cluster members (magenta), foreground perturbers (pink) and background perturbers (blue). The circles radii are proportional to the galaxy's luminosity, relative to the BCG (biggest circle) luminosity.}
   \label{fig:Mock2_Clgaldistrib}
\end{figure}

\begin{figure}[htbp]
   \centering
   \includegraphics[width=1.0\columnwidth]{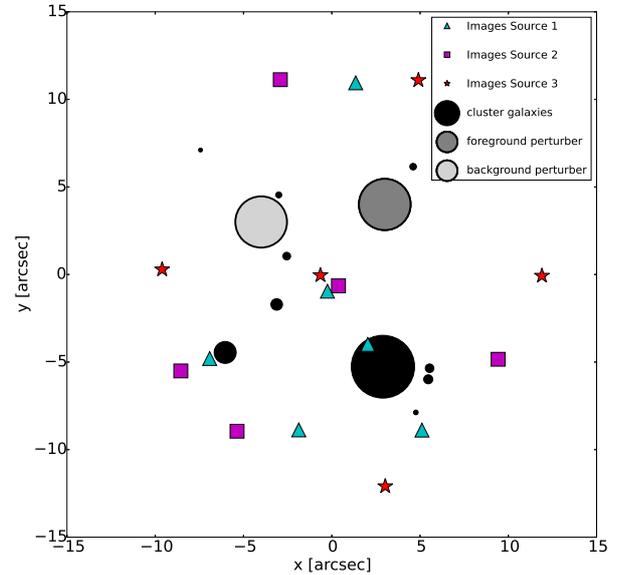} 
   \caption{Mock cluster lensing mass distribution 2. The black circles represent the lenses (cluster members), the grey circles the foreground galaxy (darker grey) and the background galaxy (lighter grey). The cyan triangles, magenta squares and red stars represent the images of the three sources, respectively at $z_{\rm s1}=1.5$, $z_{\rm s2}=2.0$, $z_{\rm s3}=2.5$. The circles radii are proportional to the galaxy's luminosity relative to the BCG (biggest circle). }
   \label{fig:Mock2_images}
\end{figure}

 We now create a more realistic mock system, composed by a cluster at $\zc=0.4$ with a halo and ten elliptical galaxies having different, realistic luminosities, axis ratios and orientations. We assume a total mass-to-light ratio corresponding to the tilt of the fundamental plane, and we scale the mass and the truncation radii according to Equation \ref{eq:ML_tiltscale}. We add two foreground perturbers at $z_{\rm fd}=0.2$, one close and one far away in projection from the cluster center, and one close-in-projection background perturber at $z_{\rm bd}=0.6$, as shown in Figure \ref{fig:Mock2_Clgaldistrib}. All the perturbers are massive (Einstein radii of $2\farcs$) and have the same Einstein radii, random ellipticity (between $0.6$ and $1$) and orientation, as shown in Table \ref{tab:Mock2_param}. We adopt a SPEMD profile for the halo, and dPIEs for the cluster members and LOS galaxies.
 We use this configuration to simulate mock lensing data, and we obtain a set of 17 multiple image positions of the 3 background sources, shown in Figure \ref{fig:Mock2_images}. We then model the parameters, i.e. all the halo parameters mentioned in Section \ref{Lensprofiles}, the cluster members' Einstein and truncation radii (with the scaling relation in Equation \ref{eq:ML_tiltscale} unless otherwise stated) and perturbers' Einstein radius (truncation radius was fixed to $15''$), with both the multi-plane set-up and the single-plane set-up (i.e. the cluster only). 
 
\subsubsection{Full multi-lens-plane modelling} 
\label{sec:Mock2MP}
 In the full multi-lens-plane model, since it has the same properties as the input (i.e. perturbers, scaling relation for cluster members etc.), we recover, within the errors, the initial parameters we have used to simulate, as shown in the MP-full column of Table \ref{tab:Mock2_param}. The modelled image positions and the magnifications are perfectly fitted, i.e. with a null total-rms offset.

 \subsubsection{Single cluster-plane modelling} 
 \label{sec:Mock2SP}
The single plane case, as shown in column SP in Table \ref{tab:Mock2_param}, shows a shift of the halo centroid position of $\sim 1\farcs4$, due to the removal of the foreground perturber which was lensing it, an overestimation of the halo Einstein radius of $\sim 3\farcs$. The halo's profile slope is less peaky in the center, and it has a core radius that is $\sim 2\farcs5$ bigger. In terms of image positions and magnification, we find that the image positions' total rms offset is $\sim 0\farcs55$, and the magnification is generally greater, up to 3 times higher.
 
 \subsubsection{Mock Cluster model 2: Assuming constant total mass-to-light ratio}
 \label{sec:Mock2constML}
\begin{figure}[htbp]
   \centering
   \includegraphics[width=\columnwidth]{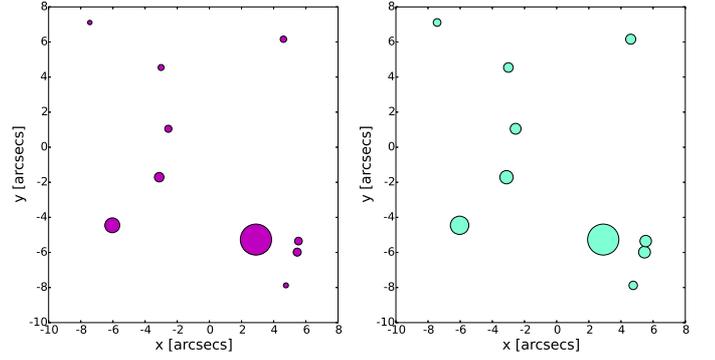} 
   \caption{Cluster galaxies' positional distribution of mock mass distribution 2, in the case of the total mass-to-light ratio equalling the tilt of the fundamental plane (left) and equalling to a constant (right). The circles radii are proportional to the galaxy's Einstein radius relative to the BCG (biggest circle) mass. }
   \label{fig:Mock2_massdistr}
\end{figure}

In this experiment, we test how the parameters change if we model assuming a different total mass-to-light relation. We choose a constant total mass-to-light ratio, which scales the Einstein and truncation radii of the cluster members as shown in Equation \ref{eq:ML_constscale}.
If we model the observables with the multi-plane system, we find, as shown in column MP-constML in Table \ref{tab:Mock2_param}, that the halo centroid position is shifted by $\sim 0.2''$. Even if the lensing effects of the foreground galaxy are still present, the halo is slightly more elliptical and the slope is slightly bigger. The cluster members have instead a bigger Einstein radius and a smaller truncation radius. All the remaining parameters are recovered within the errors. Modelling with the single-plane (SP-constML in Table \ref{tab:Mock2_param}) instead changes the halo center by $\sim 1\farcs5$, the halo Einstein radius by $\sim 2\farcs5$, which is less than what we recover in the single-plane with the previous mass-to-light ratio assumption, and underestimates the cluster galaxies masses. This model also predicts a shallower halo.\\
In terms of image positions, we find that the total rms for the multi-plane system is $\sim 0\farcs12$ and the magnification is generally smaller, within a ratio of $\sim 0.6-0.8$, showing that this approximation works slightly worse than the original total mass-to-light ratio used to simulate. The single-plane system, instead, has a total rms of $\sim 0\farcs5$, which appears to be less than what we recover using as total mass-to-light the tilt of the fundamental plane. This shows that, in the single-plane case, this approximation works slightly better than the standard SP. This is also confirmed by the magnification, which is still higher, but only up to 2 times higher. This unexpected behaviour might be explained by the fact that, as shown in Figure \ref{fig:Mock2_massdistr}, assuming a constant total mass-to-light ratio implies, given the same Einstein radius of the reference member, bigger Einstein radii for the other members (radius of circles in Figure \ref{fig:Mock2_massdistr} proportional to Einstein radius). This is compensated by a decrease of the halo Einstein radius, i.e. closer to its input value, to preserve the total mass within the total Einstein radius of the cluster. Probably a closer Einstein radius value for the halo is more efficient in reproducing the overall effect on image positions and magnification.\\

\subsubsection{Mock Cluster model 2: Assuming spherical galaxies}
\label{sec:Mock2sphere}
 The next test is to model all the galaxies (members and LOS galaxies) as if they were spherical. The multi-plane system modelling, as shown in column MP-s in Table \ref{tab:Mock2_param}, shows an offset of the halo centroid position of $\sim 0\farcs7$, and an underestimation of its Einstein radius of $\sim 3''$. The cluster galaxies and the background perturber have a bigger Einstein radii, while the foreground close-in-projection perturber has a smaller Einstein radii than the original parameter used to simulate. The image position rms is close to the observational uncertainties of $\sim 0\farcs065$, and the magnification is generally smaller, within a factor of $0.7-1$. This shows that, despite the low rms, assuming spherical galaxies creates quite a substantial offset in centroid position and Einstein radius. This suggests we should be cautious in interpreting the reconstructed cluster mass distribution, since having small rms, even comparable to positional uncertainty, does not guarantee unbiased recovery of the lens parameters. However, since our model is very simplistic (i.e. few cluster members and very massive perturbers along the line-of-sight), we suspect these results might not be so prominent in more realistic cases where there are many more galaxies and the ellipticity averages out. In the single-plane case, we have that the parameters are consistent, within the errors, to the case with elliptical galaxies and the total rms is $\sim 0\farcs55$. 

\subsubsection{Mock Cluster model 2: Assuming spherical cluster members}
\label{sec:Mock2sphclg}
To further investigate the spherical assumption on galaxies, we produce a multi-plane model similar to that done previously, but assuming only the cluster members to be spherical. Therefore, the perturbers maintain their original ellipticity used to simulate. As shown in column MP-sm in Table \ref{tab:Mock2_param}, this test still produces an offset in the centroid position, but by only $\sim 0\farcs4$; it underestimates the halo Einstein radius, but by $4\sigma$, and overestimates the cluster member Einstein radius by $\sim 3\sigma$, which is, instead, greater than the case where all the galaxies were spherical. This model also still overestimates the background perturturber and the far-in-projection foreground perturber mass, but by a smaller amount, and does recover the close-in-projection foreground mass. Unlike previously, the halo slope is underestimated, which makes the profile smoother.
The image position rms is close to the observational uncertainties of $\sim 0\farcs083$, and the magnification is generally smaller, within a factor of $0.7-0.9$. As already mention in Section \ref{sec:Mock2sphere}, we suspect these results will be mitigated in more realistic clusters cases.

\subsubsection{Mock Cluster model 2: Average surface mass density}
\label{sec:Mock2ASMD}
\begin{figure*}[htbp]
 \centering
   \includegraphics[width=0.6\textwidth]{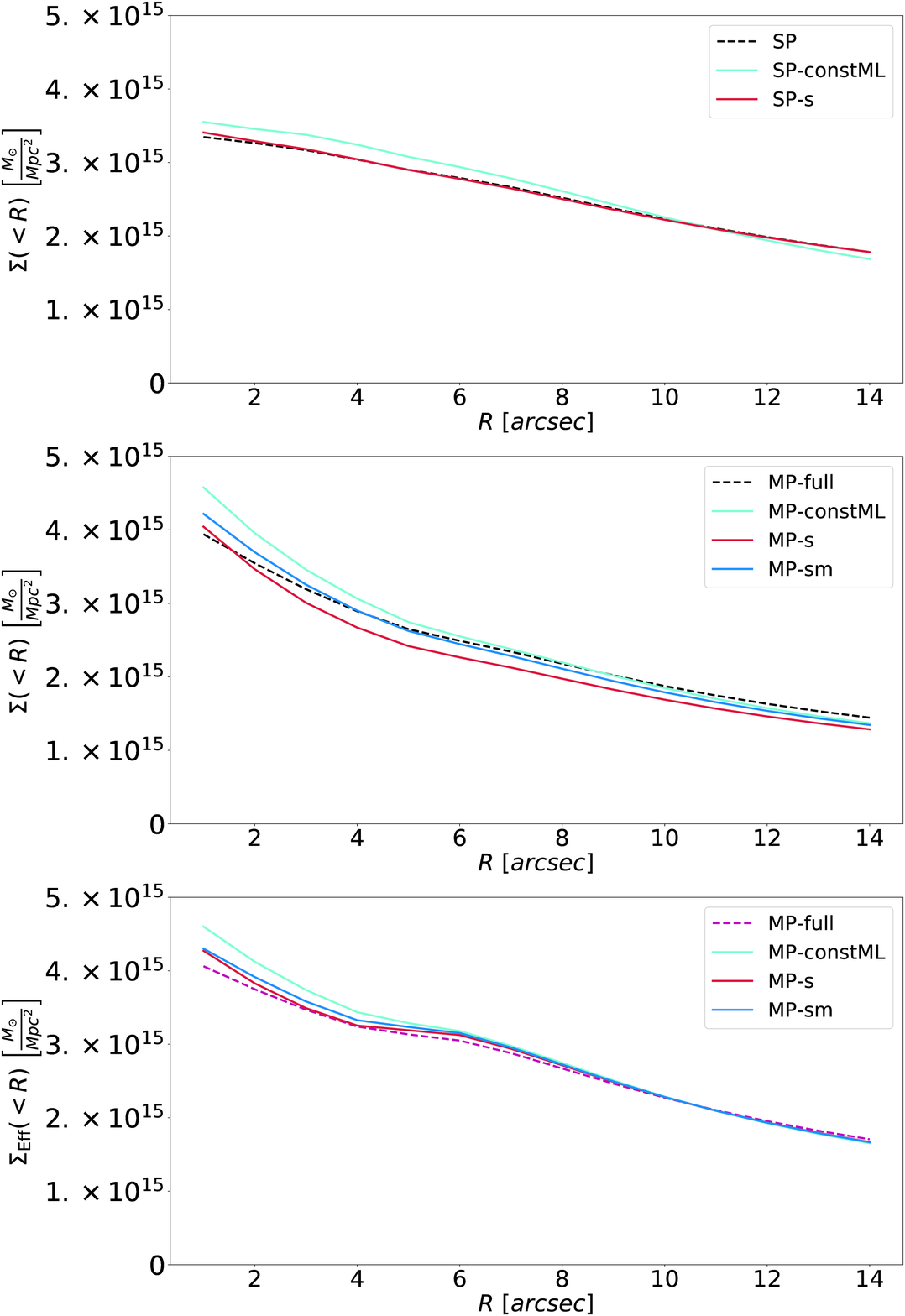} 
   \caption{Average surface mass density $\Sigma(<R)$ for source at $z_{\rm s}=2$ as a function of radius from the BCG for the best-fit models of mock mass distribution 2. From top to bottom: single plane cluster, cluster (without line-of-sight perturbers) from multi-plane models, and total multi-plane configuration. In each plot, the lines represent the model (dashed black), the model where we assumed constant mass-to-light ratio (light blue), the model where we assumed spherical galaxies (red) and, only in the multi-plane cases, the model where we assumed only spherical cluster members (blue). Note that in the total multi-plane configuration the $\Sigma_{\rm Eff}$ is relative to the total deflection angle, as explained in Section \ref{sec:Mock2ASMD}. We find that at $\theta_{\rm E,tot} \sim 10\farcs$ all models converge to a certain value of $\Sigma(<R)$ and $\Sigma_{\rm Eff}$ in the top and bottom panels, showing that strong lensing provides accurate mass enclosed within the Einstein radius.}
   \label{fig:ASMDtot}
\end{figure*}

\begin{figure*}[htbp]
   \centering
   \includegraphics[width=0.8\textwidth]{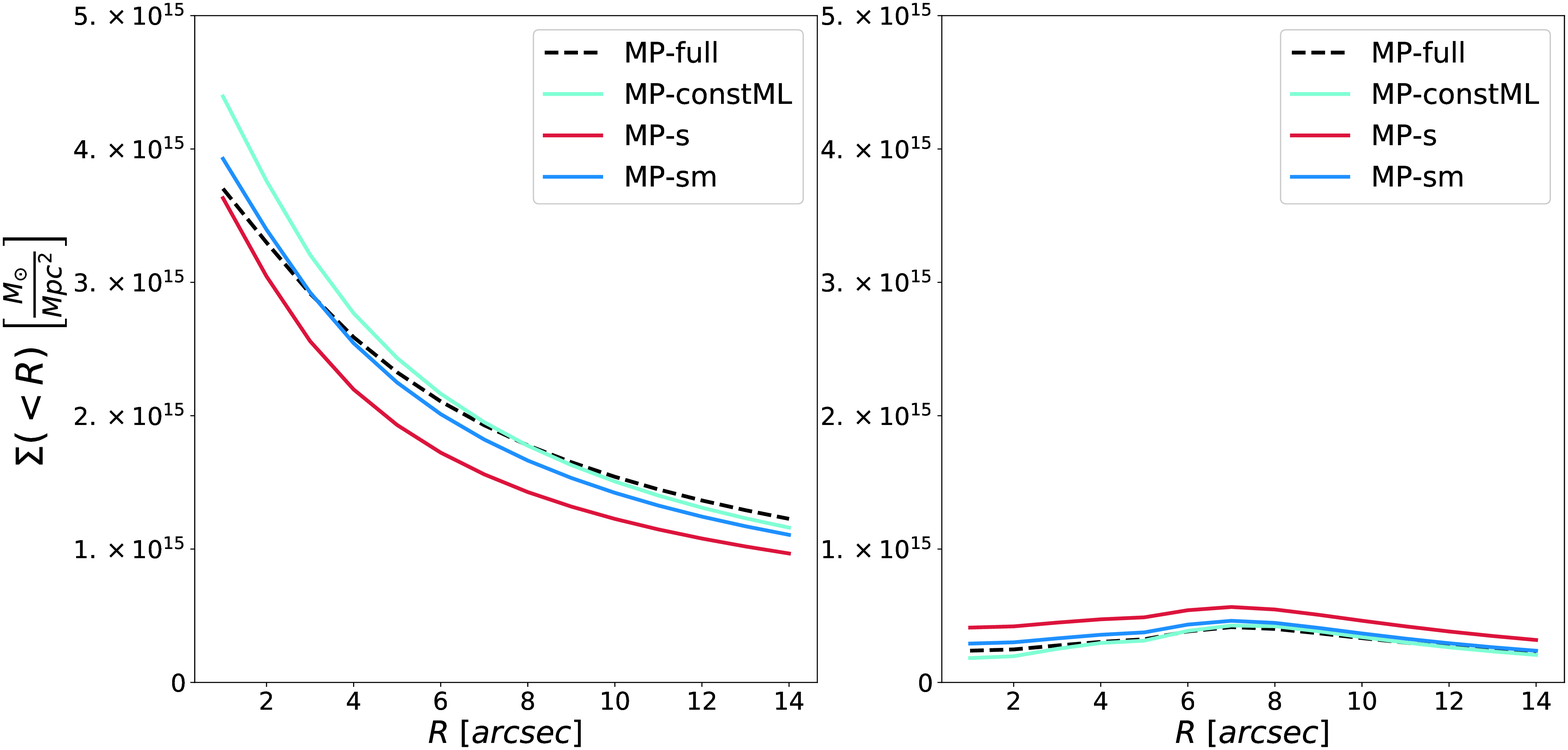} 
   \caption{Average surface mass density $\Sigma(<R)$ for source at $z_{\rm s}=2$ as a function of radius from the BCG for the multi-plane best-fit models of mock mass distribution 2. Left: cluster halo only.  Right: cluster members only. In each plot, the four lines represent the full multi-plane model (dashed black), the model where we assumed constant mass-to-light ratio (light blue), the model where we assumed spherical galaxies (red), and the model where we assumed only spherical cluster members (blue).}
   \label{fig:ASMD_MP}
\end{figure*}

\begin{figure*}[htbp]
   \centering
   \includegraphics[width=0.8\textwidth]{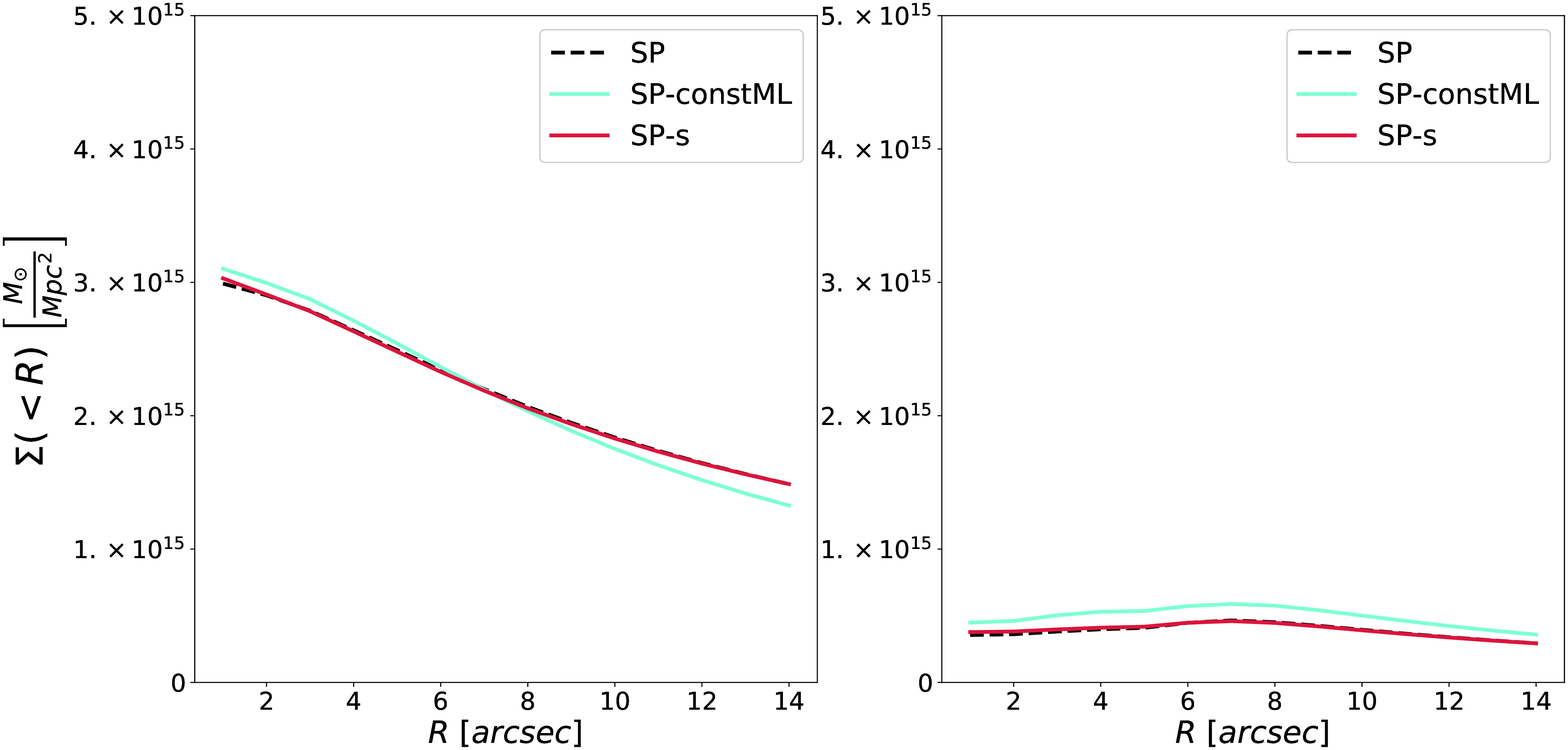} 
   \caption{Average surface mass density $\Sigma(<R)$ for source at $z_{\rm s}=2$ as a function of radius from the BCG for the single plane best-fit models of mock mass distribution 2. Left: cluster halo only.  Right: cluster members only. In each plot, the three lines represent the full multi-plane model (dashed black), the model where we assumed constant mass-to-light ratio (light blue), and the model where we assumed spherical galaxies (red).}
   \label{fig:ASMD_SP}
\end{figure*}

To explore the real differences in mass among the models, we compute the average surface mass density $\Sigma(<R)$. We obtain the convergence $\kappa $ from GLEE and multiply it by the critical surface mass density $\Sigma_{\rm crit}$ using the definition of convergence,

\be
\label{eq:kappa}
\scalebox{1.3}{$\kappa = \frac{\Sigma}{\Sigma_{\rm crit}}, $}
\ee
where $\Sigma_{\rm crit}$ is

\be
\label{eq:sigmacrit}
\scalebox{1.3}{$\Sigma_{\rm crit}=\frac{c^2}{4 \pi G} \frac{D_{\rm s}}{D_{\rm d} D_{\rm ds}}. $}
\ee
We use the cluster redshift and the median source at redshift z=2 for computing $D_{\rm d}$, $D_{\rm s}$ and $D_{\rm ds}$.
We then compute the average surface mass density, namely

\be
\label{eq:ASMD}
\scalebox{1.3}{$\Sigma (<R) = \frac{\int_{0}^{R} \! \Sigma(R') 2 \pi R'  \, \mathrm{d}R' } {\pi R^2},$}
\ee
for all the best-fit models, both multi-plane and single plane. In the multi-plane case, these quantities need to be properly defined using the multi-plane formalism we described in Section \ref{sec:MP formalism}. The deflection angle which we obtain the multi-plane kappa from is the $\alpha_{\rm tot}$ expressed by Equation \ref{eq:MPalphatot}. The quantity that comes from differentiating this $\alpha_{\rm tot}$ is what we call the $``$effective" convergence $\kappa_{\rm Eff}$. We then multiply this quantity by $\Sigma_{\rm crit}$ to obtain the $``$effective" average surface mass density $\Sigma_{\rm Eff}(<R)$. Therefore, this quantity is not a physical surface density, but the Laplacian of the total deflection angle $\alpha_ {\rm tot}$, which receives contribution from all the planes.\\
Figure \ref{fig:ASMDtot} shows $\Sigma (<R)$, from top to bottom, for both the single plane and multi-plane cluster models. Note that, instead, in the bottom panel of Figure \ref{fig:ASMDtot}, we plot the $``$effective" average surface mass density $\Sigma_{\rm Eff}(<R)$, since it contains contributions from both the cluster and the perturbers, which are at different redshifts.  The central panel of Figure \ref{fig:ASMDtot} shows, instead, the real $\Sigma(<R)$ of the cluster for the multi-plane model, which we obtain by the multi-plane best-fit models, but removing the LOS perturbers in computing the convergence $\kappa$. We see that the different models in the top and bottom panels of Figure \ref{fig:ASMDtot} tend to agree very well at $\sim 10''$, which is around the Einstein radius of the system. In the central panel of Figure \ref{fig:ASMDtot}, however, the model with the spherical galaxies assumption differs by $\sim 10\% $ from the one with elliptical galaxies. This might be due to the fact that, for that multi-plane model, we computed  $\kappa$ for the cluster only, i.e. removed some mass inside the Einstein radius, which is the quantity that lensing constraints tightly. This is also confirmed by Figure \ref{fig:ASMD_MP}, where we see that the spherical assumption is always offset from the input model (which has elliptical galaxies and mass-to-light ratio equal to the tilt of fundamental plane). Moreover, it appears that the spherical model gives more mass to the perturbers, as shown already in Table \ref{tab:Mock2_param}. If we look at the case with only the cluster members to be spherical, we see that the offset to the model is smaller with respect to the case where also the perturbers are assumed spherical, as also clear from Figure \ref{fig:ASMD_SP}, where we see that this is valid also in the single-plane case. This might be due to the fact that, leaving to the perturbers their original ellipticity gives them less mass, as we can see from the parameters. Therefore the total cluster mass is recovered better. 
Finally, we observe that the multi-plane models are generally peakier in the center, as compared to single-plane ones (see Figure \ref{fig:ASMD_SP}), since, as seen from Table \ref{tab:Mock2_param}, their core radius is smaller than that of the single plane case. 

\subsubsection{Mock Cluster model 2: Generic effects of LOS perturbers in the toy model}
\label{sec:Mock2_geneffects}
We find that the halo orientation and ellipticity are robust parameters, as we also notice in Mock cluster mass distribution 1 (discussed in Appendix), as they mainly stay, within the errors, close to the original values used to simulate.We also see that constraining the truncation radius of the galaxies is extremely difficult, as can be see also in Mock cluster mass distribution 1 (discussed in Appendix). However, in general this parameter does not appear correlated to the other parameters, and it is sampled as a flat probability distribution. We also note that the mass of the far away perturber is a quantity that is not generally correlated to the other parameters, except for a slight correlation to the cluster galaxies mass and the halo mass and slope. Moreover, its posterior probability distribution is flat, therefore this quantity is mostly unconstrained. 
In the multi-plane models we find a strong correlation between the core radius and the slope of the halo, which might be due to the fact that if we increase the core radius, the slope of the halo becomes less peaky, and that the background mass is strongly correlated with the mass of the foreground and with the halo mass. In all the single-plane cases, we find slight degeneracies between the halo ellipticity, mass and the cluster galaxies' mass. These degeneracies, as mentioned in Section \ref{sec:Mock1SP}, would keep approximately the same total mass enclosed within the multiple images.
Moreover, in Mock cluster mass distribution 1 (discussed in Appendix) we explore the effect of the single perturber, i.e. only foreground and only background. We find that including the foreground is more important for a precise image reconstruction and for recovering the input parameter of the cluster, due to the lensing effect of foreground galaxies on the cluster itself.

\begin{table*}[ht]
\caption{Constraints on lens parameters for different models of Mock cluster lensing mass distribution 2. The first column refers to the values used to simulate, the other columns refer to different models, in order, the full multi-plane, the single cluster-plane, the full multi-plane and the single cluster-plane with constant mass-to-light ratio, the multi-plane and the single cluster-plane with spherical galaxies, and finally the full multi-plane with spherical cluster members only. The values are the medians of the posterior probability distributions of the lens parameters together with their 1$\sigma$ uncertainties. The orientation is measured counter clockwise from positive x-axis.  }             
\label{tab:Mock2_param}      
\centering                          
\renewcommand{\arraystretch}{2.4}  
\resizebox{\textwidth}{!}{
\begin{tabular}{lcccccccc}        
\hline                 
Parameters & Input & MP-full    & SP    & MP-constML & SP-constML & MP-s & SP-s & MP-sm \\    

\hline                        
\hline
 $\phantom{ }\theta_{\rm E,\ fd1}$        $[\arcsec]$  & $  \phantom{-}2.00$ & $  \phantom{-}1.99_{- 0.08}^{+ 0.09}$ &  $ \phantom{-} {-}$ & $  \phantom{-}1.98_{- 0.12}^{+ 0.12}$  &  $ \phantom{-} {-}$ & $  \phantom{-}2.38_{- 0.08}^{+ 0.08}$  & $ \phantom{-} {-}$ & $  \phantom{-}1.97_{- 0.07}^{+ 0.07}$\\
 $\phantom{ }\theta_{\rm E,\ fd2}$            $[\arcsec]$  & $  2.00$ & $  2.02_{- 0.19}^{+0.19}$ &  $ \phantom{-} {-}$ & $  \phantom{-}3.10_{- 2.00}^{+ 1.4}$  &  $ \phantom{-} {-}$ & $  \phantom{-}3.10_{- 1.9}^{+ 1.4}$  & $ \phantom{-} {-}$ & $  \phantom{-}3.00_{- 1.7}^{+ 1.4}$\\
  \hline 
 $\phantom{ } x_{\rm halo}$        $[\arcsec]$ & $   0.00$ & $   0.00_{- 0.08}^{+ 0.08}$ &  $ \phantom{-}0.45_{-0.13}^{+0.13}$ &  $ \phantom{-}0.20_{-0.16}^{+0.16}$  &  $ \phantom{-}0.63_{-0.13}^{+0.12}$ &  $ \phantom{-}0.63_{-0.09}^{+0.08}$ &  $ \phantom{-}0.53_{-0.14}^{+0.11}$   &  $ \phantom{-}0.43_{-0.07}^{+0.07}$        \\
 $\phantom{ } y_{\rm halo}$        $[\arcsec]$ & $   0.00$ & $   0.00_{- 0.08}^{+ 0.09}$ &  $ 1.30_{-0.20}^{+0.20}$ &  $ \phantom{-} 0.01_{-0.09}^{+0.09}$  &  $ \phantom{-}1.30_{-0.19}^{+0.19}$ &  $ \phantom{-} -0.31_{-0.04}^{+0.05}$ &  $ \phantom{-}1.10_{-0.17}^{+0.18}$   &  $ \phantom{-} -0.10_{-0.05}^{+0.05}$ \\
 $\phantom{ } \frac{b}{a}_{\rm halo}$        & $   \phantom{-}0.80$         & $   \phantom{-}0.80_{- 0.00}^{+ 0.00}$ &  $ \phantom{-}0.86_{-0.04}^{+0.03}$  &  $ \phantom{-}0.75_{-0.02}^{+0.02}$ &  $ \phantom{-}0.84_{-0.04}^{+0.03}$ &  $ \phantom{-} 0.71_{-0.03}^{+0.02}$ &  $ \phantom{-}0.85_{-0.05}^{+0.04}$   &  $ \phantom{-} 0.74_{-0.02}^{+0.02}$\\
 $\phantom{ }\theta_{\rm halo}$  $[rad]$ & $  0.00$  & $  0.00_{-0.00}^{+ 0.00}$ &  $ 0.00_{-0.02}^{+0.02}$ &  $ 0.00_{-0.05}^{+0.05}$ &  $ 0.00_{-0.02}^{+0.02}$ &  $ 0.02\pi_{-0.00}^{+0.00}$  &  $ 0.00_{-0.02}^{+0.02}$ &  $ 0.01\pi_{-0.00}^{+0.00}$ \\
 $\phantom{ } \theta_{\rm E,\ halo}$     $[\arcsec]$  & $   \phantom{-} 10.00$ & $   \phantom{-} 9.99_{- 0.16}^{+ 0.16}$ &  $ \phantom{-} 12.90_{-0.79}^{+1.00}$ &  $ \phantom{-} 9.83_{-0.40}^{+0.38}$ &  $ \phantom{-} 12.50_{-0.72}^{+0.93}$ &  $ \phantom{-} 6.87_{-0.39}^{+0.47}$ &  $ \phantom{-} 12.90_{-1.00}^{+2.20}$ &  $ \phantom{-} 8.81_{-0.28}^{+0.27}$  \\
 $\phantom{ } r_{\rm c,\ halo}$        $[\arcsec]$ & $  \phantom{-} 2.00$ & $  \phantom{-} 2.00_{- 0.23}^{+ 0.24}$ &  $\phantom{-} 4.50_{-0.74}^{+0.77}$ &  $\phantom{-}1.80_{-0.28}^{+0.32}$ &  $\phantom{-}4.90_{-0.72}^{+0.71}$ &  $\phantom{-}1.10_{-0.12}^{+0.14}$ &  $\phantom{-}4.10_{-1.30}^{+0.77}$ &  $\phantom{-}1.81_{-0.17}^{+0.17}$  \\
 $\phantom{ }\gamma_{\rm halo}$     & $   0.40$    & $   0.40_{- 0.02}^{+ 0.02}$ &  $ 0.56_{-0.11}^{+0.13}$ &  $ 0.46_{-0.04}^{+0.04}$ &  $ 0.68_{-0.13}^{+0.09}$ &  $ 0.39_{-0.02}^{+0.02}$ &  $ 0.52_{-0.15}^{+0.10}$ &  $ 0.44_{-0.02}^{+0.02}$    \\
 $\phantom{ }\theta_{\rm E,\ g}$      $[\arcsec]$      & $   \phantom{-}3.00$     & $   \phantom{-}3.00_{- 0.09}^{+ 0.09}$ &  $ \phantom{-}2.20_{-0.35}^{+0.38}$ &  $ \phantom{-}3.65_{-0.23}^{+0.25}$ &  $ \phantom{-}2.20_{-0.33}^{+0.34}$ &  $ \phantom{-}3.27_{-0.12}^{+0.12}$ &  $ \phantom{-}2.10_{-0.47}^{+0.43}$ &  $ \phantom{-}3.40_{-0.12}^{+0.12}$   \\
 $\phantom{ } r_{\rm t,\ g}$  $[\arcsec]$ & $  15.0$  & $  15.10_{-1.00}^{+1.10}$ &  $ 220.03_{-34.23}^{+82.84}$ & $  8.50_{-0.97}^{+1.15}$ & $  69.60_{-32.68}^{+51.62}$ & $  34.00_{-8.51}^{+11.94}$ & $  506.70_{-407.89}^{+146.0}$ & $  14.50_{-1.33}^{+1.62}$    \\
    \hline 
 $\phantom{ }\theta_{\rm E,\ bd}$     $[\arcsec]$ & $  \phantom{-}2.00$ & $  \phantom{-}2.00_{- 0.25}^{+ 0.24}$ &  $ \phantom{-} {-}$ & $  \phantom{-}2.30_{- 0.42}^{+ 0.41}$  &  $ \phantom{-} {-}$ & $  \phantom{-}3.48_{- 0.23}^{+ 0.20}$  & $ \phantom{-} {-}$ & $  \phantom{-}3.01_{- 0.21}^{+ 0.22}$\\
     \hline
     \hline
 $\phantom{ } \rm rms$     $[\arcsec]$ &  $ \phantom{-} $ &  $ \phantom{-} 4.09 \times 10^{-5}$ & $  \phantom{-} 0.549$  &  $ \phantom{-} 0.126$ & $  \phantom{-} 0.511$  & $ \phantom{-}0.069$ & $  \phantom{-} 0.547$ & $  \phantom{-} 0.083$ \\  
    
\hline                                   
\end{tabular}}
\end{table*}

\section{\ourlens\ mass distribution}
\label{sec:0416}

\subsection{Observation of \ourlens}
\label{sec:MACSintro}
\begin{figure*}[htbp]
   \centering
   \includegraphics[width=\textwidth]{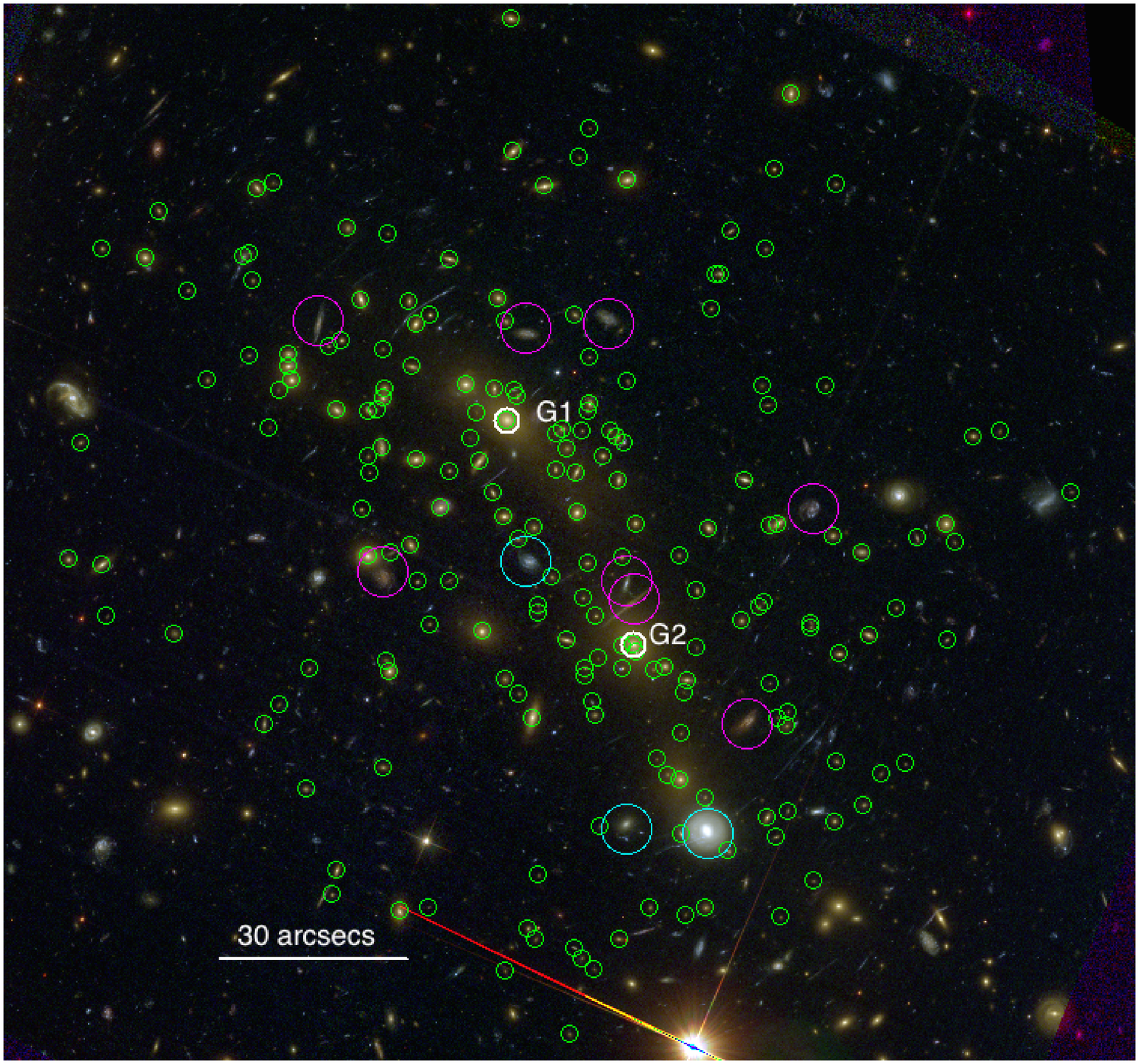} 
   \caption{Colour image of \ourlens\ obtained through a combination of the HST/ACS and WFC3 filters. We mark the selected 11 secondary lenses that we included in our model to account for the LOS contribution, using different colors for foreground (cyan) and background (magenta). In green we circle the cluster members. In white are marked the two BCGs. North is up and East is left.}
   \label{fig:0416}
\end{figure*}

In this section we model the mass distribution of the HFF cluster \ourlens\ (from here on, \macs), which is a massive luminous cluster at $z_{\rm c}=0.396$, with a very large spectroscopic data set along the line of sight \citep{Balestra2016}. This cluster, shown in Figure \ref{fig:0416}, was first discovered within the MAssive Cluster Survey (MACS) in the X-rays by \citet{Mann&Ebeling2012}. It shows an elongation in the NE-SW direction and was firstly identified as a merger. Since its discovery, \macs\ has been extensively studied \citep[e.g.][]{Zitrin2013,Jauzac2014,Jauzac2015,Grillo2015,Caminha2016,Hoag2016,Kawamata2016,Natarajan2017,Bonamigo2017} as it represents an efficient lens for magnifying sources and producing multiple images (strong lensing regime), as shown in Figure \ref{fig:0416images}. In most recent studies \citep[][]{Grillo2015,Caminha2016} the image position was reconstructed with extremely high precision ($\Delta_{\rm rms}\sim 0\farcs3$ in \citet{Grillo2015}, with a set of 30 spectroscopically identified images, and  $\Delta_{\rm rms}\sim 0\farcs59$  in \citet{Caminha2016}, with a set of 107 spectroscopically confirmed images), reaching the limit of what can be achieved by neglecting LOS contribution in the modelling. 
To study the effect of the LOS galaxies, we create a mock mass distribution by building a model of \macs\ that reproduces the observables as close as possible, and we use the model parameters to produce simulated observables. We then model the mock \macs\ with different assumptions and compare them to the input mock model to assess the impact of LOS perturbers, as was done in previous sections.

\subsection{\ourlens\ best-fit model}
\label{sec:m0416_bestfit}
\begin{figure*}[htbp]
   \centering
   \includegraphics[width=\textwidth]{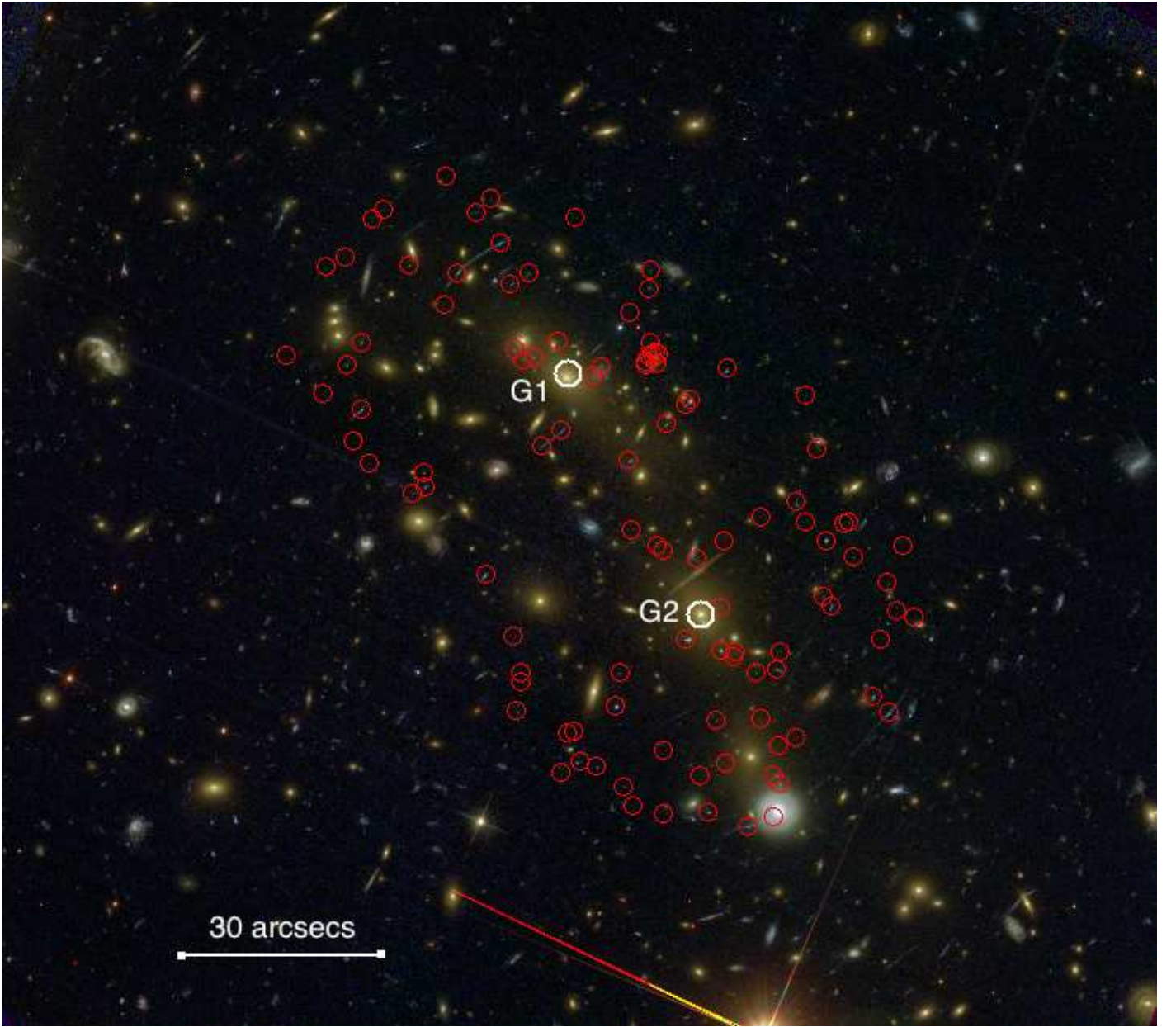} 
   \caption{Colour image of \ourlens\ obtained through a combination of the HST/ACS and WFC3 filters. We mark the selected 107 images we included in our model, corresponding to 37 sources that range from redshift $\sim1$ to $\sim6$ . In white are marked the two BCGs. North is up and East is left.}
   \label{fig:0416images}
\end{figure*}

\begin{figure}[htbp]
   \centering
  \includegraphics[width=\columnwidth]{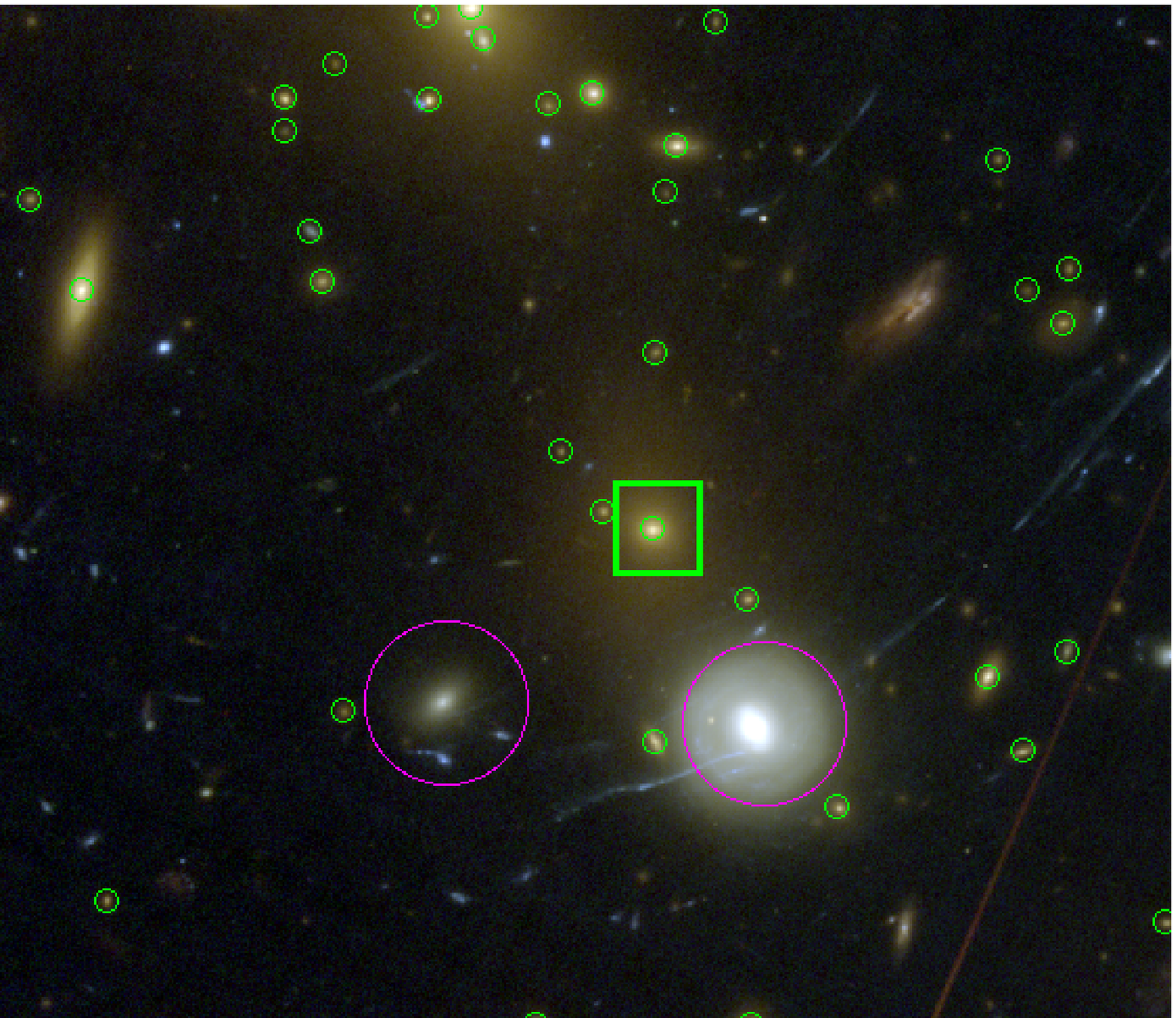} 
   \caption{Colour image of \ourlens\ obtained through a combination of the HST/ACS and WFC3 filters. Green circles mark the cluster members. We mark the massive member (in green square) that we do not scale using scaling relations, since it is very close to the two foreground galaxies (marked in magenta) and its observed magnitude might be affected by their light contamination.}
   \label{fig:0416pecmember}
\end{figure}

We model the mass distribution of \macs\ with a single-plane setup very similar to that presented by \citet{Caminha2016}, but optimised independently with GLEE. This model consists of 193 cluster members, 3 dark matter halos and it is modelled on a set of 107 images (shown in Figure \ref{fig:0416images}) corresponding to 37 sources, all of them spectroscopically confirmed. We then include into the model 11 chosen foreground and background galaxies, which are listed in Table \ref{tab:Perturbers} and discussed in Section \ref{sec:MACSlos}. We use truncated dual pseudo-isothermal elliptical mass distributions (dPIEs) to represent the galaxies (members and perturbers), and pseudo-isothermal elliptical mass distribution (PIEMDs) for the halos, as these profiles were shown to reproduce better the observables \citep{Grillo2015}. We assume all the galaxies to be spherical. Two halos are located on the North-East (halo 1), South-West (halo 2) and centered, respectively, on the Northern and Southern BCGs. A third smaller halo is located to the Eastern part of the northern halo.\\
To obtain the best-fit model that includes the perturbers, we model the halo parameters (centroid position ($x_{\rm h}, y_{\rm h}$), axis ratio $\frac{b}{a}$, orientation $\theta$, Einstein radius $\theta_{\rm E,h}$, core radius $r_{\rm c,h}$), the Einstein radii and truncation radii of the cluster members and perturbers. We scale the Einstein radii and truncation radii of the cluster members using the tilt of fundamental plane for the total mass-to-light ratio (Equation \ref{eq:ML_tiltscaleproof}), as done in Section \ref{sec:Mock2input}, using Equation \ref{eq:ML_tiltscale}. This mass-to-light ratio was shown to better reproduce the image position of this cluster by \citet{Grillo2015}. We do not include in the scaling relation of the cluster members a luminous galaxy which is very close to the bright foreground galaxy (and also to a less massive foreground galaxy) in the South-West region of the cluster (see Figure \ref{fig:0416pecmember}). The reason for this choice is that the light contamination from the foreground galaxy in that region does not allow an accurate estimation of the cluster member's magnitude, which is also affected by the magnification due to the foreground galaxies. We quantified the same magnification effect due to the foreground galaxies on other close-by cluster members, but we saw it was negligible. We find that allowing that particular member to vary freely (both its Einstein radius, and truncation radius) decreases substantially the $\chi^2$ of around $\sim18\%$.
We scale the Einstein radius and truncation radius of the perturbers as explained in Section \ref{sec:MACSlos}. We use an error on the images which corresponds to the observational uncertainty $\sim0\farcs06$ (one pixel). However, we use a special treatment for images with high magnification forming arcs, since they have a more elliptical shape. We therefore introduce elliptical errors for those systems, with the minor axis of around one pixel, the major axis between $0\farcs2-0\farcs4$ (depending on the spatial extent of the arc) and oriented along the direction of the arc.

\subsubsection{LOS secondary lenses}
\label{sec:MACSlos}
The secondary lenses are chosen among the brightest objects of the HST image, selected using near-IR (F160W) luminosities. We expect these brightest LOS galaxies to have the most significant effects on the modelling, compared to other fainter LOS galaxies. For the background perturber we take into account the magnification effect due to the cluster lensing. We keep the number of perturbers relatively low ($ 11$) for computational efficiency and such that the addition of cosmic voids in the environment of the lens is not necessary. Figure \ref{fig:0416} shows the cluster \macs\  with the perturbers, indicated by color depending on whether they are foreground or background galaxies. To be noted that the brighest foreground perturber in the southern region of the cluster was already included in previous models \citep{Johnson2014, Richard2014, Grillo2015, Caminha2016}, but at the cluster redshift. We include this foreground bright galaxy to its actual redshift z=0.1124 (first object in Table \ref{tab:Perturbers}). Each of the 11 LOS perturber has two additional parameters (Einstein and truncation radii), and we scale the Einstein and truncation radii of these LOS galaxies with respect to the brighter ones. We scale separately the foreground and background, using the definition of apparent magnitude and correcting the observed magnitude of the background for the magnification effect due to the presence of the foreground and the cluster, as follows:
\be
\label{eq:magnitude}
\scalebox{1.2}{$ m- m _{\rm ref} = -2.5\ {\rm log}_{\rm 10} \left( \frac{F}{F_{\rm ref}} \right), $}
\ee
where the flux F
\be
\label{eq:flux}
\scalebox{1.2}{$ F = \frac{L}{4 \pi D_{\rm l} ^2},$}
\ee
with $ D_{\rm l}$ the luminosity distance to the object. Substituting in the definition of magnitude,
\be
\label{eq:magdiff}
\scalebox{1.2}{$ m- m _{\rm ref} = -2.5\ {\rm log}_{\rm 10} \left( \frac{\frac{L}{4 \pi D_{\rm l} ^2}}{\frac{L _{\rm ref}}{4 \pi D_{\rm l, ref}^2}} \right), $}
\ee
we obtain
\be
\label{eq:magdiff2}
\scalebox{1.2}{$ m- m _{\rm ref} = -2.5\ {\rm log}_{\rm 10} \left( \frac{L}{L_{\rm ref}} \left( \frac{D_{\rm l, ref}}{D_{\rm l}}\right)^2 \right).$}
\ee
If we now want to obtain the luminosity, we revert the equation to
\be
\label{eq:L_ratio}
\scalebox{1.2}{$ \frac{L}{L_{\rm ref}} = 10^{\frac{\rm m_{\rm ref} -m}{2.5}} \left( \frac{D_{\rm l}}{D_{\rm l, ref}}\right)^2,  $}
\ee
and we scale it with the magnification $\mu$,
\be
\label{eq:L_scaled}
\scalebox{1.2}{$ L_{\rm r}= \frac{\rm \mu_{\rm ref}}{\rm \mu} \frac{L}{L_{\rm ref}} = \frac{\rm \mu_{\rm ref}}{\rm \mu} 10^{\frac{\rm m_{\rm ref} -m}{2.5}} \left( \frac{D_{\rm l}}{D_{\rm l, ref}}\right)^2 , $}
\ee
where the magnification value for each perturber galaxy is obtained from the multi-plane best-fit model with only the foreground galaxies and the cluster. 
Therefore, the scaling is obtained by multiplying the reference quantity to the luminosity ratio $L_{\rm r}$ as shown in Equation \ref{eq:ML_tiltscale}. We let the bright foreground galaxy at $z=0.1124$ free to vary, and we scale the remaining two foreground galaxies with respect to the galaxy at $z_{\rm f, ref}=0.1126$, and the 8 background with respect to that at $z_{\rm b, ref}=0.5004$. The reference perturbers were chosen since they have the greater apparent magnitude among, respectively, the foreground and background LOS galaxies, as shown in Table \ref{tab:Perturbers}.

\begin{table*}[ht]
\caption{Selected secondary lenses added to the \ourlens\  cluster, reported in redshift order, from the lowest to the highest. The first two columns report the RA-DEC coordinates, the third column the ID, the fourth column the redshift, the fifth the quality flag  of redshift (where a value of 3 represents ``secure”), and the last three the R, F814W and F160W magnitudes. The line separates the galaxies that are on the foreground and on the background of the cluster.}             
\label{tab:Perturbers}      
\centering                          
\renewcommand{\arraystretch}{2.4}  
\begin{tabular}{lccccccc}        
 \hline                 
 $ \rm RAJ2000$  & $\rm DEJ2000$ & $\rm ID$  & $\rm z$ & $\rm QF$ & $\rm R $ & $\rm F814W$ & $\rm F160W$
 \\    
\hline                        
 \hline
 $64.0285$ & $-24.0857$ & ${\rm CLASHVLTJ041606.8-240508}$ & $0.1124$ & $3$ & $-$ & $16.70$ & $16.09$\\
 $64.0323$ & $-24.0854$ & ${\rm CLASHVLTJ041607.7-240507}$ & $0.1126$ & $3$ & $20.66$ & $20.76$ & $20.25$\\
 $64.0370$ & $-24.0738$ & ${\rm CLASHVLTJ041608.9-240425}$ & $0.1533$ & $3$ & $21.25$ & $21.17$ & $21.03$\\
 \hline
 $64.0323$ & $-24.0747$ & ${\rm CLASHVLTJ041607.8-240429}$ & $0.4678$ & $3$ & $21.05$ & $21.49$ & $19.92$\\
 $64.0331$ & $-24.0631$ & ${\rm IDVLTJ041607.9-240347 } $ &  $0.4848$ & $3$ & $-$ & $22.46$ & $20.99$\\
 $64.0322$ & $-24.076 $ & ${\rm IDVLTJ041607.7-240433} $ &   $0.5004$  & $3$ & $-$ & $20.47$ & $19.49$\\
 $64.0473$ & $-24.0633$ & ${\rm CLASHVLTJ041611.3-240347}$ & $0.5277$ & $3$ & $21.22$ & $21.44$ & $20.80$\\
 $64.0372 $ & $-24.0637$ & ${\rm CLASHVLTJ041608.9-240349}$ & $0.5376$ & $3$ & $21.63$ & $21.41$ & $20.76$\\
 $64.0441$ & $-24.0745$ & ${\rm CLASHVLTJ041610.6-240427}$ & $0.7093$ & $3$ & $21.32$ & $21.03$ & $19.96$\\
 $64.0233$ & $-24.0714$ & ${\rm CLASHVLTJ041605.6-240417}$ & $0.7358$ & $3$ & $21.58$ & $21.33$ & $21.42$\\
 $64.0265$ & $-24.0809$ & ${\rm CLASHVLTJ041606.4-240451}$ & $0.7362$ & $3$ & $21.20$ & $20.70$ & $19.79$\\

 \hline 
\end{tabular}
\end{table*}

\subsubsection{Results of best-fit model of \ourlens}
\label{sec:ASMD_bestfit}

\begin{figure*}[htbp]
   \centering
   \includegraphics[width=0.9\textwidth]{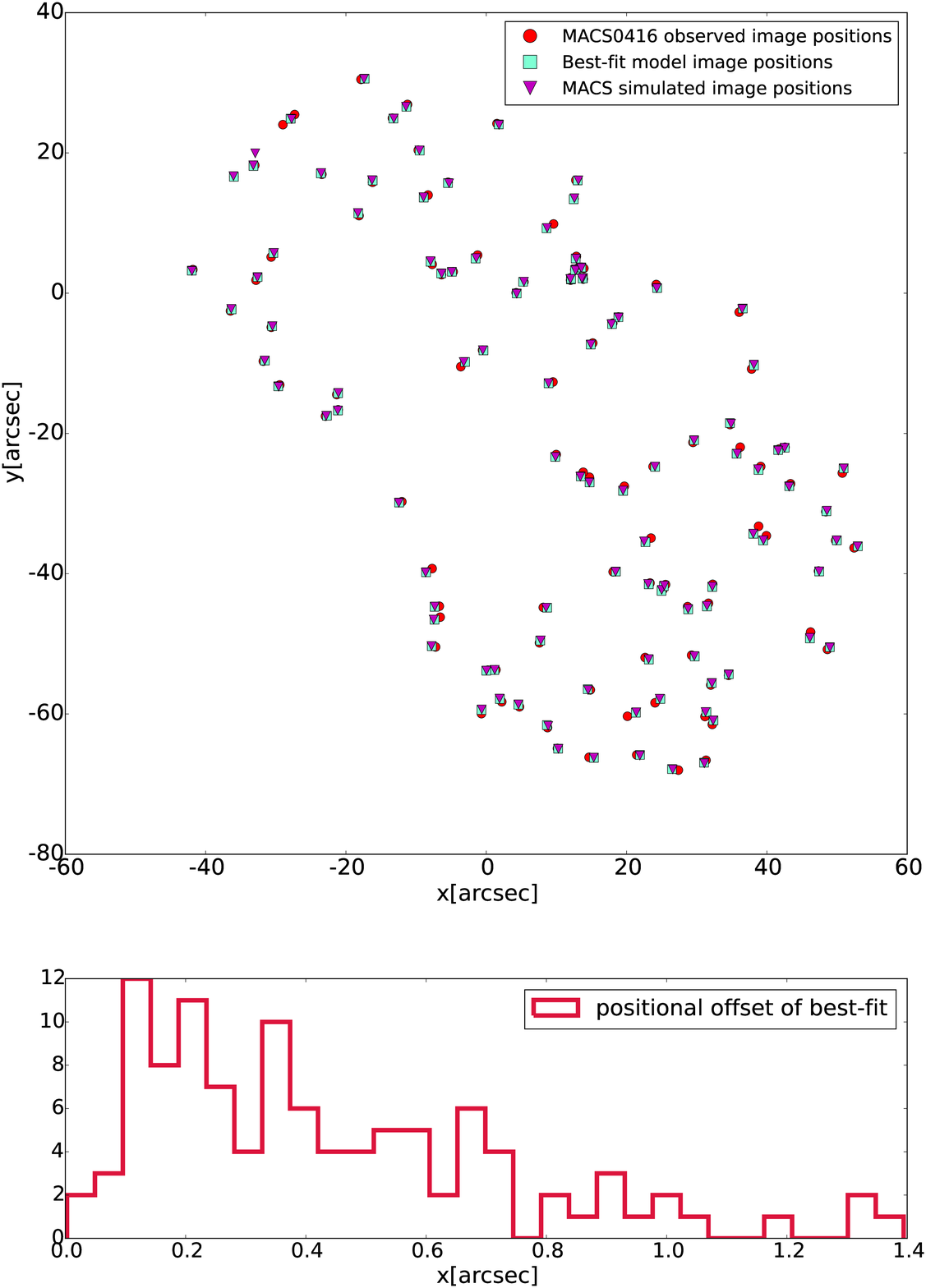} 
   \caption{The upper panel shows the predicted image positions (squares) of \ourlens\ and the simulated image positions (triangles) of our MACS model (with introduced gaussian scatter), in comparison to the observed image positions (circles) of \ourlens\ . The lower panel shows the histogram of the positional offsets between the observed and the model-predicted image positions. The rms for our best-fit model of \ourlens\ is $\sim0\farcs53$.  }
   \label{fig:MACSimages}
\end{figure*}

\begin{figure}[htbp]
   \centering
   \includegraphics[width=\columnwidth]{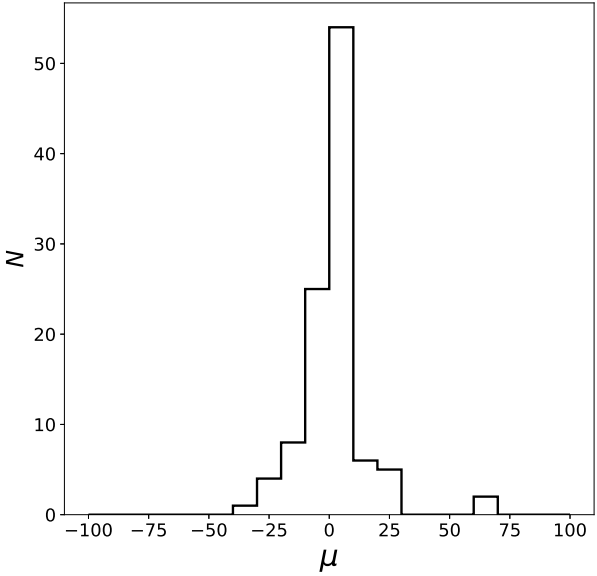} 
   \caption{Predicted magnification histogram for the best-fit model of \ourlens. There is one image with predicted magnification of $\sim4\times10^{4}$ (not included in our plot for visualization convenience), which is one of the arc-shaped images in the North-East region ($-20\farcs,20\farcs$) of \ourlens, as shown in Figure \ref{fig:MACSimages}.}
   \label{fig:m0416_magn}
\end{figure}

\begin{figure*}[htbp]
   \centering
   \includegraphics[width=0.9\textwidth]{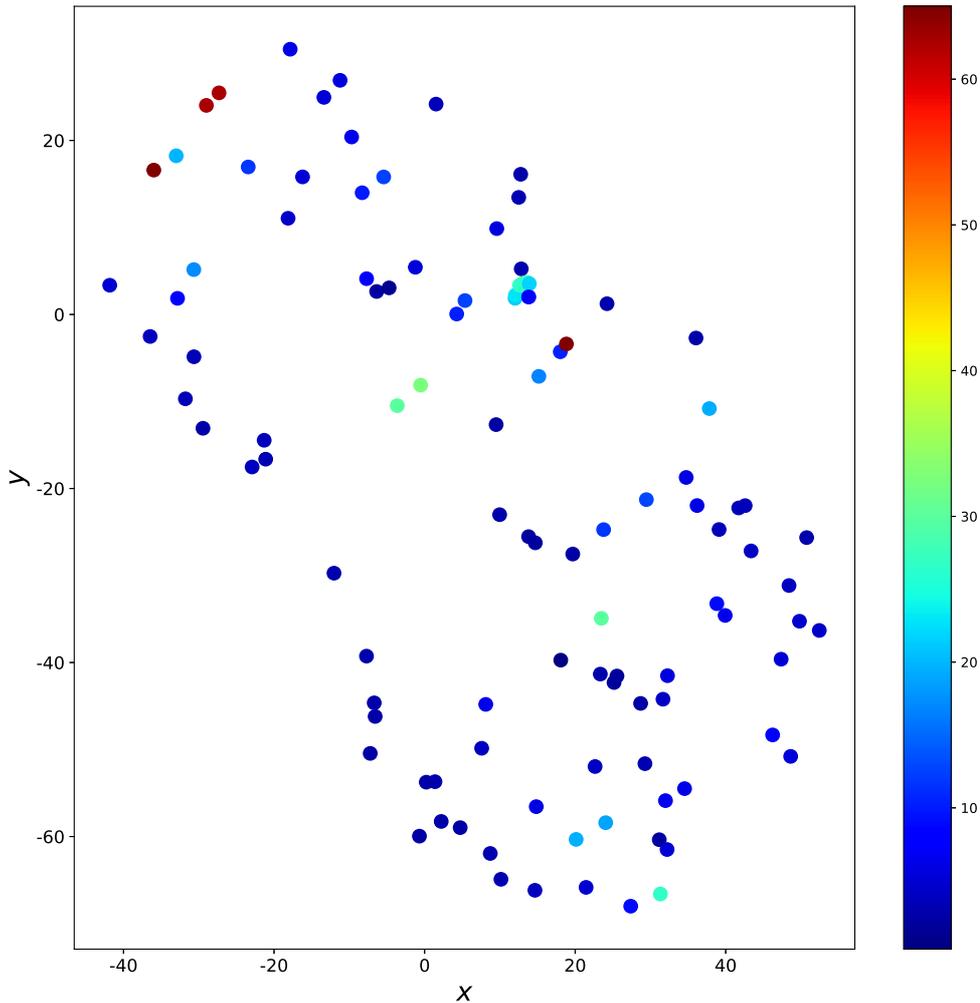} 
   \caption{Magnification for the image positions predicted by best-fit model of \ourlens. There is one image with predicted magnification of $\sim4\times10^{4}$ and one with predicted magnification of $\sim600$ whose magnification value was set to the border value of $\mu =65$ for visualization convenience. The two images are in the systems with high magnification and whose positional uncertainty was treated as elliptical due to their arc-like shape.}
   \label{fig:m0416_magn_pos}
\end{figure*}

\begin{figure*}[htbp]
   \centering
   \includegraphics[width=0.9\textwidth]{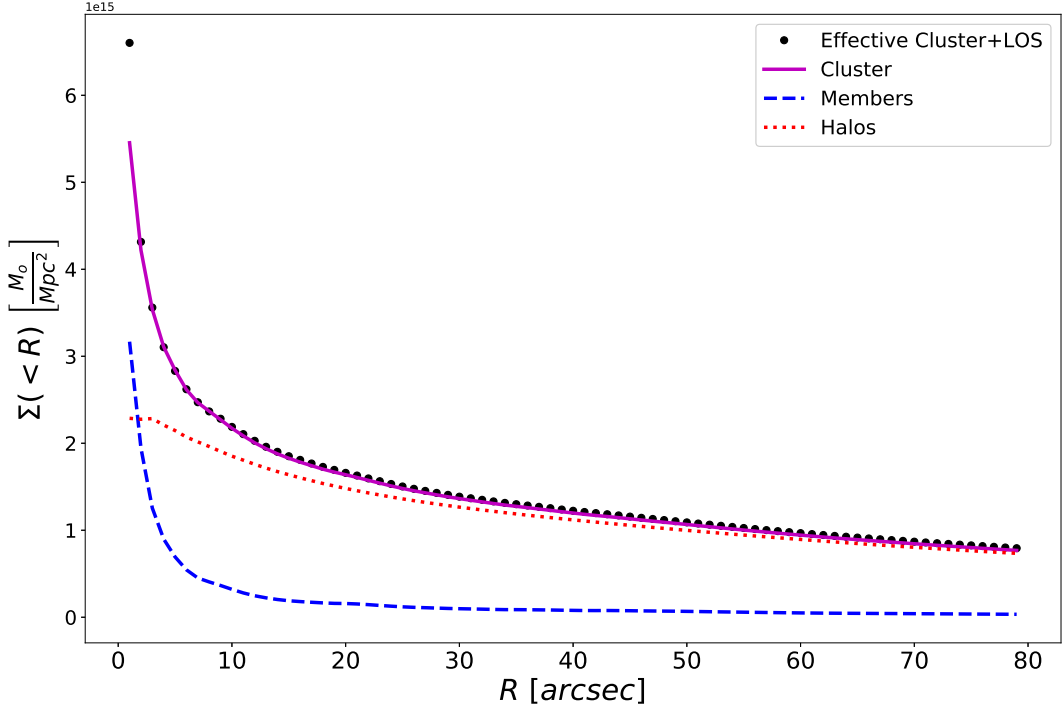} 
   \caption{Average surface mass density $\Sigma (<R)$ (for source at redshift $z_{\rm s}=3$) as a function of radius (from G1) for the best-fit model of \ourlens\ for the total mass of the cluster (magenta line), for the cluster members (dashed blue) and for the dark matter halos (dotted red). The black points represents the effective average surface mass density $\Sigma_{\rm eff}(<R)$ of the cluster and the LOS galaxies, as explained in Section \ref{sec:Mock2ASMD}. We look at the $\Sigma(<R)$ within $\sim80''$, which approximately corresponds to the HST FoV at z=0.396 ($\sim420\ \rm kpc$). }
   \label{fig:m0416_asmd}
\end{figure*}

\begin{figure*}[htbp]
   \centering
   \includegraphics[width=\textwidth]{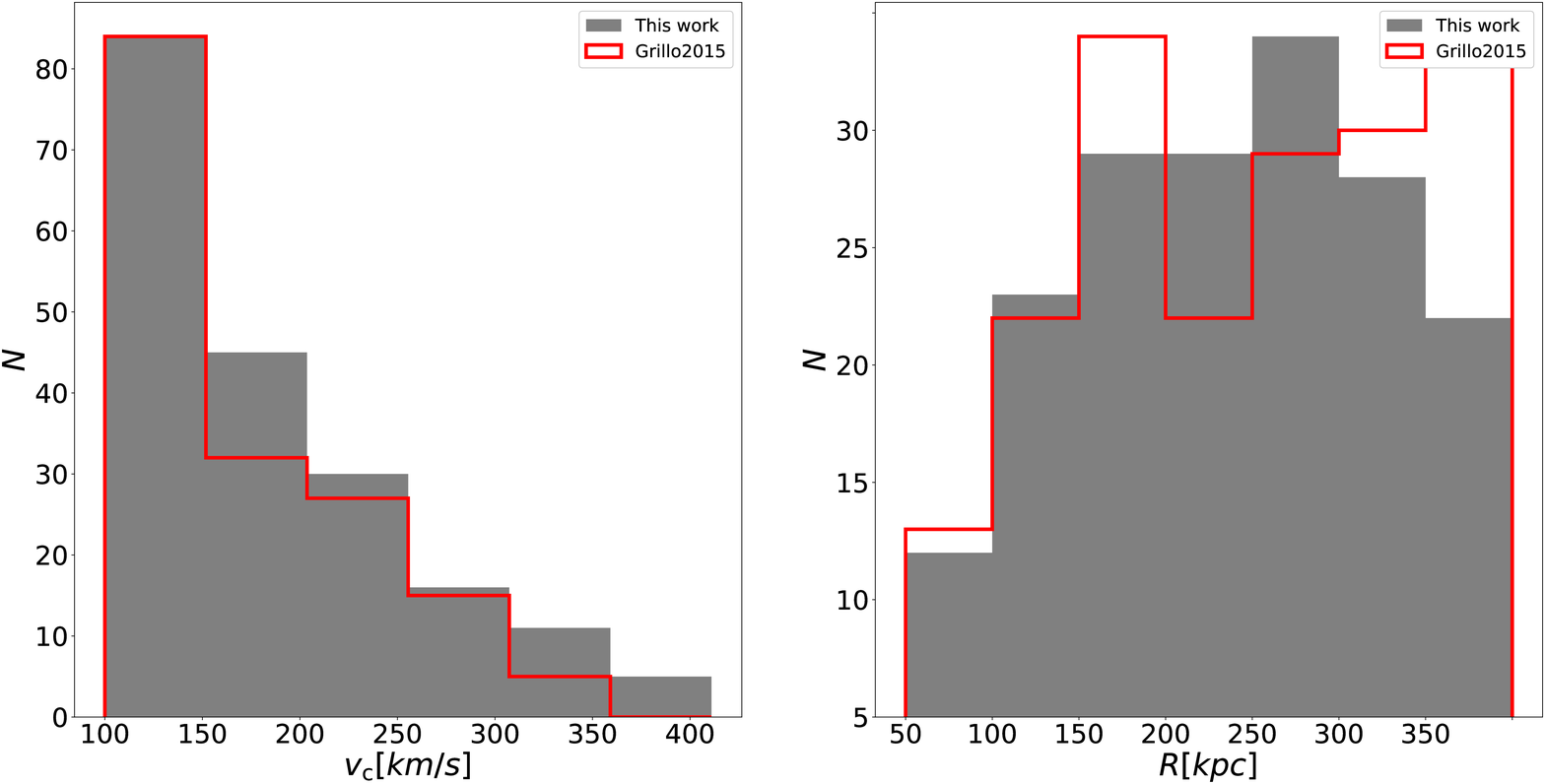} 
   \caption{Number of subhalos as a function of the projected distance from the cluster lens center (right) and as a function of circular velocity (left) for our bestfit model of \ourlens\ within an aperture of $\sim400\ \rm kpc$.}
   \label{fig:m0416_Nsigma}
\end{figure*}

Our best-fit model can reproduce the image positions with an $\rm rms\sim0\farcs53$  (see Figure \ref{fig:MACSimages}), i.e. approximately 8-9 pixels. Previous works with the same set of images \citep[]{Caminha2016} was able to reproduce the observed image positions with a rms of $\sim0\farcs59$, therefore our model with the addition of the LOS galaxies shows an improvement of $\sim0\farcs06$. Our model has a $\chi^2 \sim7.5\times10^3$. 
The best-fit mass distribution parameters are shown in Table \ref{tab:bestfit_param}. We remark that $1''$ at the redshift of the first foreground galaxy $z_{\rm fd}\sim0.1 $ corresponds to $\sim1.8\ {\rm kpc}$, while at the cluster redshift $\zc =0.396$, $ 1''$ corresponds to $\sim5.34\ \rm kpc$. Therefore, the truncation radius of the bright foreground galaxy is $ \sim300\ \rm kpc $, which is a typical value for massive galaxies. However, we find from the posterior probabilities that the truncation radii of the LOS galaxies are mostly unconstrained due to the fact that they are isolated galaxies. The truncation radii of the cluster members are instead smaller, due to the tidal stripping of their dark matter halos.
Figure \ref{fig:MACSimages} shows the observed image positions of \ourlens\ and the predicted image positions of our best-fit model, which are in very good agreement, as can also be seen by the histogram of their positional offset. However, our model predicts 106 images of 107 observed, since two images in the North-East region of the cluster are predicted as a unique one (see Figure \ref{fig:MACSimages})\footnote{These images belong to a source at redshift $z_{\rm s}=3.2387$, that has three arc-shape images with high magnification ($>100$). Two of them are very close to each other and they are predicted by our model to lie on the same side of the critical curves.}.
The predicted magnification of the 106 images is shown in Figure \ref{fig:m0416_magn} and \ref{fig:m0416_magn_pos}. One image has a very high predicted magnification ($\sim4\times10^{4}$), that is not shown in the plot for visualization convenience\footnote{This image belong to the system of arcs in the North-East part of the cluster mentioned previously ($z_{\rm s}=3.2387$). Again, this shows that this system, being close to critical curves, is very highly magnified.}.
   In Figure \ref{fig:m0416_asmd}, we plot the average surface mass density for the best-fit model of \ourlens. We find that neglecting the LOS contribution does not affect the total $\Sigma$ reconstruction significantly (compare black points and magenta lines on Figure \ref{fig:m0416_asmd}). We also find that most of the contribution on the outskirts (from $\sim20''$ from the Northern BCG), is due to the halo mass, while in the very center ($\sim5''$ from the Northern BCG), the contribution of the galaxies is more prominent. We have used the unlensed positions (i.e. those they wold have if the perturbers were not there) for the halos and members when computing the average surface mass density of the cluster, halos and members separately. We also study the velocity dispersion of the cluster galaxies, that we compute from the Einstein radius of a singular isothermal sphere (SIS), which is expressed by
\be
\label{eq:theta_E_SIS}
\scalebox{1.2}{$\theta_{\rm E} = 4\pi \frac{\sigma^2_{\rm v}}{c^2}.$}
\ee
for source at infinite redshift. We then look (Figure \ref{fig:m0416_Nsigma}) at the number of galaxies with a certain velocity dispersion, therefore mass, and we confirm the trend already pointed out by \citet{Grillo2015}, that the amount of small mass halos seems to be in better agreement with simulations, whereas we find more high mass subhalos than predicted by cosmological simulations \citep[details on cosmological simulations can be found in][]{Bonafede2011,Contini2012,Grillo2015}. Regarding the radial distribution of substructures, we compute the position of the cluster members by removing the lensing effect of the foreground galaxies, and we compute the radial distance from the barycenter $R_{\rm b}= (1\farcs21,-6\farcs85)$, which was obtained by the weighted average

\be
\label{eq:barycenter}
\scalebox{1.3}{$R_{\rm b} = \frac{\int \! \Sigma(R')  R'  \, \mathrm{d}R' } {\int \! \Sigma(R')  \, \mathrm{d}R' }.$}
\ee
We find our model to slightly underpredict the number of substructures at small radii and at large radii ($\sim300-400\ {\rm kpc}$), while to overpredict at radii $\sim200-300\ {\rm kpc}$ with respect to the model of \citet{Grillo2015}. This implies a better agreement with cosmological simulations at smaller radii and at large radii, whereas a worse agreement for radii $\sim200-300\ {\rm kpc}$. \\

\begin{table}[ht]
\caption{Best-fit lens parameters for our model of \ourlens\ mass distribution (including 3 dark matter halos, 193 cluster members, 3 foreground galaxies and 8 background galaxies).  The first halo (`$\rm h1$') is the one located in the northern part of the cluster, the second halo (`$\rm h2$') in the southern one, the third (`$\rm h3$') is the smaller northern halo. The subscripts `$\rm fd$' and `$\rm bd$' refers, respectively, to the parameters of the foreground and background galaxies. The subscripts `$\rm g$' and `$\rm gp$' refers to the parameters of the reference cluster galaxy and the peculiar member we allow to vary out of the scaling relations. The orientation is measured counter clockwise from positive x-axis. The center of the coordinates is the position of the Northern BCG (G1). }             
\label{tab:bestfit_param}      
\centering                          
\renewcommand{\arraystretch}{1.4}  
\resizebox{\width}{!}{
\begin{tabular}{lcccccccccccccccccc}        
\hline                 
Parameters & Best-fit parameter values  \\    

\hline                        
\hline
 $\phantom{ }\theta_{\rm E,\ fd1}$        $[\arcsec]$ & $  \phantom{-}1.7$ \\
 $\phantom{ } r_{\rm t,\ fd1}$            $[\arcsec]$   & $  148.58$ \\
  $\phantom{ }\theta_{\rm E,\ fd2}$            $[\arcsec]$   & $  0.01 $ \\
   $\phantom{ } r_{\rm t,\ fd2}$            $[\arcsec]$   & $  99.71$ \\
  \hline 
 $\phantom{ } x_{\rm h1}$        $[\arcsec]$ & $   -1.51 $ \\
 $\phantom{ } y_{\rm h1}$        $[\arcsec]$ & $   0.51$ \\
 $\phantom{ } \frac{b}{a}_{\rm h1}$                 & $   \phantom{-}0.34$\\
 $\phantom{ }\theta_{\rm h1}$  $[rad]$   & $  2.50$ \\ 
 $\phantom{ } \theta_{\rm E,\ h1}$     $[\arcsec]$ & $   \phantom{-} 16.66$ \\
 $\phantom{ } r_{\rm c,\ h1}$        $[\arcsec]$ & $  \phantom{-} 7.92$ \\
  $\phantom{ } x_{\rm h2}$        $[\arcsec]$ & $   20.73$ \\
 $\phantom{ } y_{\rm h2}$        $[\arcsec]$ & $   -38.72$ \\
 $\phantom{ } \frac{b}{a}_{\rm h2}$                 & $   \phantom{-}0.39$  \\
 $\phantom{ }\theta_{\rm h2}$  $[rad]$   & $  2.19$ \\ 
 $\phantom{ } \theta_{\rm E,\ h2}$     $[\arcsec]$ & $   \phantom{-} 31.26$ \\
 $\phantom{ } r_{\rm c,\ h2}$        $[\arcsec]$ & $  \phantom{-} 12.52$ \\ 
 $\phantom{ } x_{\rm h3}$        $[\arcsec]$ & $   -33.73$ \\
 $\phantom{ } y_{\rm h3}$        $[\arcsec]$ & $   10.52$ \\
 $\phantom{ } \frac{b}{a}_{\rm h3}$                 & $   \phantom{-}0.90$ \\
 $\phantom{ }\theta_{\rm h3}$  $[rad]$   & $  2.56$ \\ 
 $\phantom{ } \theta_{\rm E,\ h3}$     $[\arcsec]$ & $   \phantom{-} 5.89$ \\ 
 $\phantom{ } r_{\rm c,\ h3}$        $[\arcsec]$ & $  \phantom{-} 8.95$ \\ 
 $\phantom{ }\theta_{\rm E,\ g}$      $[\arcsec]$           & $   \phantom{-}1.64$ \\ 
 $\phantom{ } r_{\rm t,\ g}$  $[\arcsec]$   & $  3.48$ \\
 $\phantom{ }\theta_{\rm E,\ gp}$      $[\arcsec]$           & $   \phantom{-}1.54$ \\ 
 $\phantom{ } r_{\rm t,\ gp}$  $[\arcsec]$   & $  4.98$ \\
     \hline
 $\phantom{ } \theta_{\rm E,\ bd}$     $[\arcsec]$ &  $ \phantom{-} 0.13$   \\  
 $\phantom{ } r_{\rm t,\ bd}$     $[\arcsec]$ &  $ \phantom{-} 33.50$   \\  
  
\hline                                   
\end{tabular}}
\end{table}

\subsection{Mock MACS input}
\label{sec:MACSinput}
Our best-fit model of \macs\ and its environment, can reproduce the image positions with a $\rm rms\sim0\farcs53$  (see Figure \ref{fig:MACSimages}), which is, as already mentioned, the current best-fit model obtained with this set of images. Our set of simulated constraints are the 106 predicted image positions from the best-fit model of \ourlens. Since one image of one system was not predicted by our model (see Section \ref{sec:ASMD_bestfit} for discussion), we decided to tweak the observed position of the third image manually in a way that its position was on the other side of the critical curve with respect to the second image and therefore such that it was predicted by our best-fit model. We then used the image position predicted by best-fit \ourlens\ model as the 107th constraint for creating the mock system that we call “Mock MACS”, making our set of constraints to a total of 107 images. As mentioned before, the images' positional uncertainties corresponds to the observational uncertainty $\sim0\farcs06$, but we use a special treatment for images with high magnification forming arcs, introducing elliptical errors for those systems, as explained in Section \ref{sec:m0416_bestfit}. To obtain the simulated image positions, we then shift the 107 predicted image positions (of the best-fit model) by a random number, drawn from a 2D-gaussian distribution with $\sigma=0\farcs06$, in both x and y direction to introduce a random scatter. In the case of highly magnified images forming arcs, we draw random numbers from a 2D elliptical Gaussian distribution, with $\sigma_1, \sigma_2$ equal to the minor and major axis of the error ellipses on those images. We then rotate the Gaussian to align with the direction of the arcs, since arcs are tangentially oriented. We use these 107 shifted image positions as our observables. The initial $\chi^2$ of the input Mock MACS cluster is 246.

\subsection{Mock MACS models}
\label{sec:MACSmodels}
Once we have simulated a set of images (107 in total, from 37 sources), we model all the halo parameters (centroid position, ellipticity, orientation, Einstein radius, core radius) and the Einstein radii and truncation radii of the cluster members and of the perturbers (see Table \ref{tab:MACS_param}), with different assumptions to assess the effect of these foreground LOS perturbers. We model the Mock MACS mass distribution using:
\begin{itemize}
\item[1.] MP-full: the full multi-plane treatment, i.e. including all the LOS perturbers at the different redshifts.  
\item[2.] SP1: assuming only the bright foreground pertuber at the cluster redshift \citep[similar to models in][]{Johnson2014, Richard2014, Grillo2015, Caminha2016}
\item[3.] SP: assuming no line-of-sight perturbers, and including only cluster members. 
\item[4.] MP-fore: multi-plane including only the three foreground galaxies.
\item[5.] MP-back: multi-plane including only the eight background galaxies. 
\item[6.] MP-constML: full multi-plane with different scaling relation, i.e. using constant mass-to-light ratio as shown in Equation \ref{eq:ML_constscale}.
\item[7.] SP-constML: single plane using constant mass-to-light ratio (Equation \ref{eq:ML_constscale}).
\end{itemize}
We describe the results of each of the models below.

\begin{table*}[ht]
\caption{Constraints on lens parameters for different models of the Mock MACS mass distribution (3 dark matter halos, 193 cluster members, 3 foreground galaxies and 8 background galaxies). The first halo (`$\rm h1$') is the one located in the northern part of the cluster, the second halo (`$\rm h2$') in the southern part, the third (`$\rm h3$') is the smaller northern halo. The subscripts `$\rm fd$' and `$\rm bd$' refer, respectively, to the parameters of the foreground and background galaxies. The subscripts `$\rm g$' and `$\rm gp$' refer to the parameters of the reference cluster galaxy and the peculiar member we allow to vary out of the scaling relations. The values are the medians of the posterior probability distributions of the lens parameters together with their 1$\sigma$ uncertainties. The orientation is measured counter clockwise from positive x-axis. The center of the coordinates is the position of the Northern BCG (G1).}             
\label{tab:MACS_param}      
\centering                          
\renewcommand{\arraystretch}{1.4}  
\resizebox{\width}{!}{
\begin{tabular}{lccccccccccccccccccc}        
\hline                 
Parameters & Input & MP-full & SP1 & SP & MP-fore & MP-back & MP-constML & SP-constML  \\    

\hline                        
\hline
 $\phantom{ }\theta_{\rm E,fd1}$        $[\arcsec]$ & $  \phantom{-}1.7$ & $  \phantom{-}1.73_{-0.04}^{+0.04} $ & $  \phantom{-}1.19_{-0.05}^{+0.05} $ & $  \phantom{-} - $ & $  \phantom{-}1.67_{-0.06}^{+0.04} $ & $  \phantom{-} - $  & $  \phantom{-}1.82_{-0.09}^{+0.07} $  & $  \phantom{-} - $ \\
 $\phantom{ } r_{\rm tfd1}$            $[\arcsec]$   & $  148.58$ & $  \phantom{-}147.9_{-9.8}^{+0.79}$ & $  \phantom{-}112.9_{-0.84}^{+0.97} $ & $  \phantom{-} - $ & $  \phantom{-}111.7_{-0.12}^{+0.18} $ & $  \phantom{-} - $  & $  \phantom{-}147.00_{-2.5}^{+2.0} $  & $  \phantom{-} - $ \\
  $\phantom{ }\theta_{\rm Ef,d2}$            $[\arcsec]$   & $  0.01 $ & $  \phantom{-}0.03_{-0.02}^{+0.03}$ & $  \phantom{-}  - $ & $  \phantom{-} - $ & $  \phantom{-} 0.02_{-0.01}^{+0.03} $ & $  \phantom{-} - $  & $  \phantom{-}0.06_{-0.03}^{+0.06} $  & $  \phantom{-} - $ \\
   $\phantom{ } r_{\rm t,fd2}$            $[\arcsec]$   & $  99.71$ & $  \phantom{-}99.34_{-3.2}^{+0.46}$ & $  \phantom{-} - $ & $  \phantom{-} - $ & $  \phantom{-} 94.64_{-0.14}^{+0.11} $ & $  \phantom{-} - $  & $  \phantom{-}59.00_{-5.70}^{+11.00} $  & $  \phantom{-} - $ \\
  \hline 
 $\phantom{ } x_{\rm h1}$        $[\arcsec]$ & $   -1.51 $ & $   -1.55_{-0.09}^{+0.09}$ & $   -1.92_{-0.17}^{+0.12} $ & $  -1.2_{-0.31}^{+0.22} $ & $  -1.42_{-0.08}^{+0.07} $ & $  -1.7_{-0.28}^{+0.35} $  & $  -1.20_{-0.15}^{+0.16} $  & $  \phantom{-}0.03_{-0.91}^{+0.89} $ \\
 $\phantom{ } y_{\rm h1}$        $[\arcsec]$ & $   0.51$ & $  \phantom{-}0.56_{-0.08}^{+0.09}$ & $  \phantom{-}1.03_{-0.10}^{+0.13} $ & $  \phantom{-} 0.30_{-0.18}^{+0.18} $ & $  \phantom{-} 0.62_{-0.05}^{+0.05} $ & $  \phantom{-} 0.48_{-0.30}^{+0.24} $ & $  \phantom{-}0.36_{-0.14}^{+0.14} $  & $  -0.36_{-0.69}^{+0.77} $ \\
 $\phantom{ } \frac{b}{a}_{\rm h1}$                 & $   \phantom{-}0.34$ & $  \phantom{-}0.34_{-0.00}^{+0.00}$ & $  \phantom{-}0.34_{-0.01}^{+0.01} $ & $  \phantom{-} 0.27_{-0.01}^{+0.01}$ & $  \phantom{-} 0.34_{-0.00}^{+0.01} $ & $  \phantom{-} 0.28_{-0.02}^{+0.01} $ & $  \phantom{-}0.34_{-0.01}^{+0.01} $  & $  \phantom{-}0.27_{-0.06}^{+0.03} $ \\
 $\phantom{ }\theta_{\rm h1}$  $[rad]$   & $  2.50$ & $  \phantom{-}2.50_{-0.004}^{+0.003}$ & $  \phantom{-}2.52_{-0.01}^{+0.01} $ & $  \phantom{-} 2.54_{-0.01}^{+0.01} $ & $  \phantom{-} 2.50_{-0.00}^{+0.00} $ & $  \phantom{-} 2.54_{-0.01}^{+0.01} $ & $  \phantom{-}2.49_{-0.01}^{+0.01} $  & $  \phantom{-}2.53_{-0.03}^{+0.03} $ \\ 
 $\phantom{ } \theta_{\rm E,h1}$     $[\arcsec]$ & $   \phantom{-} 16.66$ & $  \phantom{-}16.68_{-0.13}^{+0.14}$ & $  \phantom{-}15.2_{-0.24}^{+0.20} $ & $  \phantom{-} 11.4_{-0.19}^{+0.35} $ & $  \phantom{-}16.47_{-0.08}^{+0.15} $ & $  \phantom{-} 13.0_{-0.71}^{+0.39} $ & $  \phantom{-}15.9_{-0.39}^{+0.30} $  & $  \phantom{-}11.00_{-2.00}^{+1.20} $ \\
 $\phantom{ } r_{\rm c,h1}$        $[\arcsec]$ & $  \phantom{-} 7.92$ & $  \phantom{-}7.92_{-0.09}^{+0.08}$ & $  \phantom{-}7.30_{-0.16}^{+0.14} $ & $  \phantom{-} 6.12_{-0.21}^{+0.29} $ & $  \phantom{-}7.92_{-0.09}^{+0.1} $ & $  \phantom{-} 6.72_{-0.42}^{+0.34} $ & $  \phantom{-}6.66_{-0.18}^{+0.22} $  & $  \phantom{-}5.00_{-1.20}^{+0.77} $ \\
  $\phantom{ } x_{\rm h2}$        $[\arcsec]$ & $   20.73$ & $  \phantom{-}20.71_{-0.07}^{+0.07}$ & $  \phantom{-}20.63_{-0.09}^{+0.10} $ & $  \phantom{-} 21.37_{-0.15}^{+0.18} $ & $  \phantom{-}20.54_{-0.06}^{+0.06} $ & $  \phantom{-} 21.5_{-0.28}^{+0.27} $ & $  \phantom{-}20.18_{-0.10}^{+0.10} $  & $  \phantom{-}20.30_{-0.49}^{+0.35} $ \\
 $\phantom{ } y_{\rm h2}$        $[\arcsec]$ & $   -38.72$ & $  -38.67_{-0.09}^{+0.11}$ & $  -38.16_{-0.16}^{+0.14} $ & $  -39.31_{-0.23}^{+0.16} $ & $  -38.34_{-0.09}^{+0.07} $ & $   -39.8_{-0.50}^{+0.45} $ & $  -38.06_{-0.19}^{+0.21} $  & $  -37.90_{-0.66}^{+0.94} $ \\
 $\phantom{ } \frac{b}{a}_{\rm h2}$                 & $   \phantom{-}0.39$  & $  \phantom{-}0.39_{-0.00}^{+0.00}$ & $  \phantom{-}0.38_{-0.00}^{+0.00} $ & $  \phantom{-} 0.34 _{-0.01}^{+0.01} $ & $  \phantom{-}0.39_{-0.00}^{+0.00} $ & $  \phantom{-} 0.34_{-0.01}^{+0.01} $ & $  \phantom{-}0.38_{-0.00}^{+0.00} $  & $  \phantom{-}0.33_{-0.01}^{+0.11} $ \\
 $\phantom{ }\theta_{\rm h2}$  $[rad]$   & $  2.19$ & $  \phantom{-}2.20_{-0.00}^{+0.00}$ & $  \phantom{-}2.19_{-0.00}^{+0.00} $ & $  \phantom{-} 2.17_{-0.00}^{+0.00} $ & $  \phantom{-}2.20_{-0.00}^{+0.00} $ & $  \phantom{-} 2.17_{-0.00}^{+0.00} $ & $  \phantom{-}2.20_{-0.0}^{+0.0} $  & $  \phantom{-}2.18_{-0.02}^{+0.01} $ \\ 
 $\phantom{ } \theta_{\rm E,h2}$     $[\arcsec]$ & $   \phantom{-} 31.26$ & $  \phantom{-}31.50_{-0.24}^{+0.33}$ & $  \phantom{-}34.89_{-0.21}^{+0.37} $ & $  \phantom{-} 39.38_{-0.27}^{+0.31} $ & $  \phantom{-}32.41_{-0.14}^{+0.08} $ & $  \phantom{-} 38.60_{-0.51}^{+0.96} $ & $  \phantom{-}31.00_{-0.48}^{+0.51} $  & $  \phantom{-}39.70_{-1.10}^{+2.00} $ \\
 $\phantom{ } r_{\rm c,h2}$        $[\arcsec]$ & $  \phantom{-} 12.52$ & $  \phantom{-}12.7_{-0.15}^{+0.14}$ & $  \phantom{-}13.93_{-0.12}^{+0.18} $ & $  \phantom{-} 16.4_{-0.18}^{+0.23} $ & $  \phantom{-}12.87_{-0.09}^{+0.08} $ & $  \phantom{-} 16.20_{-0.29}^{+0.44} $ & $  \phantom{-}12.10_{-0.18}^{+0.21} $  & $  \phantom{-}15.60_{-0.62}^{+0.84} $ \\ 
 $\phantom{ } x_{\rm h3}$        $[\arcsec]$ & $   -33.73$ & $  -33.74_{-0.18}^{+0.15}$ & $  -33.27_{-0.32}^{+0.25} $ & $   -33.88_{-0.45}^{+0.33} $ & $  -33.81_{-0.09}^{+0.19} $ & $   -35.60_{-0.83}^{+1.20} $ & $  -34.10_{-0.42}^{+0.41} $  & $  -32.80_{-3.20}^{+2.60} $ \\
 $\phantom{ } y_{\rm h3}$        $[\arcsec]$ & $   10.52$ & $  \phantom{-}10.60_{-0.12}^{+0.12}$ & $  \phantom{-}10.20_{-0.17}^{+0.18} $ & $  \phantom{-} 12.00_{-0.42}^{+0.23} $ & $  \phantom{-}10.67_{-0.17}^{+0.08} $ & $  \phantom{-} 11.80_{-1.00}^{+0.46} $ & $  \phantom{-}10.7_{-0.23}^{+0.24} $  & $  \phantom{-}11.60_{-1.10}^{+1.30} $ \\
 $\phantom{ } \frac{b}{a}_{\rm h3}$                 & $   \phantom{-}0.90$ & $  \phantom{-}0.97_{-0.06}^{+0.02}$ & $  \phantom{-}0.96_{-0.04}^{+0.03} $ & $  \phantom{-} 0.59_{-0.06}^{+0.05} $ & $  \phantom{-}0.92_{-0.02}^{+0.02} $ & $  \phantom{-} 0.74_{-0.09}^{+0.06} $ & $  \phantom{-}0.96_{-0.05}^{+0.03} $  & $  \phantom{-}0.61_{-0.17}^{+0.15} $ \\
 $\phantom{ }\theta_{\rm h3}$  $[rad]$   & $  2.56$ & $  \phantom{-}2.53_{-0.21}^{+0.13}$ & $  \phantom{-}2.62_{-0.18}^{+0.18} $ & $  \phantom{-} 2.83_{-0.06}^{+0.08} $ & $  \phantom{-}2.57_{-0.07}^{+0.05} $ & $  \phantom{-} 2.97_{-0.12}^{+0.18} $ & $  \phantom{-}2.30_{-0.55}^{+0.32} $  & $  \phantom{-}2.86_{-0.15}^{+0.29} $ \\ 
 $\phantom{ } \theta_{\rm E,h3}$     $[\arcsec]$ & $   \phantom{-} 5.89$ & $  \phantom{-}5.60_{-0.22}^{+0.29}$ & $  \phantom{-}4.95_{-0.21}^{+0.16} $ & $  \phantom{-} 7.34_{-0.33}^{+0.22} $ & $  \phantom{-}5.39_{-0.07}^{+0.07} $ & $  \phantom{-} 6.14_{-0.46}^{+0.51} $ & $  \phantom{-}6.14_{-0.32}^{+0.41} $  & $  \phantom{-}7.30_{-1.30}^{+2.20} $ \\ 
 $\phantom{ } r_{\rm c,h3}$        $[\arcsec]$ & $  \phantom{-} 8.95$ & $  \phantom{-}8.40_{-0.39}^{+0.53}$ & $  \phantom{-}7.12_{-0.30}^{+0.39} $ & $  \phantom{-} 12.8_{-0.49}^{+0.37} $ & $  \phantom{-}7.93_{-0.10}^{+0.07} $ & $  \phantom{-} 10.8_{-0.72}^{+1.10} $ & $  \phantom{-}8.73_{-0.58}^{+0.65} $  & $  \phantom{-}11.00_{-1.90}^{+3.70} $ \\ 
 $\phantom{ }\theta_{\rm E,m}$      $[\arcsec]$           & $   \phantom{-}1.64$ & $  \phantom{-}1.57_{-0.03}^{+0.05}$ & $  \phantom{-}1.63_{-0.04}^{+0.05} $ & $  \phantom{-} 1.79_{-0.07}^{+0.07} $ & $  \phantom{-}1.60_{-0.03}^{+0.04} $ & $  \phantom{-} 1.77_{-0.07}^{+0.10} $ & $  \phantom{-}1.52_{-0.03}^{+0.03} $  & $  \phantom{-}1.90_{-0.15}^{+0.17} $ \\ 
 $\phantom{ } r_{\rm t,m}$  $[\arcsec]$   & $  3.48$ & $  \phantom{-}3.70_{-0.22}^{+0.17}$ & $  \phantom{-}3.50_{-0.18}^{+0.16} $ & $  \phantom{-} 4.66_{-0.34}^{+0.21} $ & $  \phantom{-}3.54_{-0.10}^{+0.11} $ & $  \phantom{-} 4.46_{-0.31}^{+0.41} $ & $  \phantom{-}3.07_{-0.05}^{+0.10} $  & $  \phantom{-}3.13_{-0.12}^{+0.46} $ \\
 $\phantom{ }\theta_{\rm E,mp}$      $[\arcsec]$           & $   \phantom{-}1.54$ & $  \phantom{-}1.58_{-0.05}^{+0.04}$ & $  \phantom{-}1.85_{-0.10}^{+0.21} $ & $  \phantom{-} 1.55_{-0.11}^{+0.11} $ & $  \phantom{-}1.60_{-0.04}^{+0.04} $ & $  \phantom{-} 1.56_{-0.12}^{+0.14} $ & $  \phantom{-}1.53_{-0.08}^{+0.07} $  & $  \phantom{-}1.50_{-0.27}^{+0.29} $ \\ 
 $\phantom{ } r_{\rm t,mp}$  $[\arcsec]$   & $  4.98$ & $  \phantom{-}4.92_{-0.15}^{+0.07}$ & $  \phantom{-}2.16_{-0.39}^{+0.22} $ & $  \phantom{-} 4.89_{-0.22}^{+0.09} $ & $  \phantom{-}4.94_{-0.09}^{+0.05} $ & $  \phantom{-} 4.68_{-0.63}^{+0.23} $ & $  \phantom{-}4.85_{-0.28}^{+0.12} $  & $  \phantom{-}4.83_{-0.95}^{+0.17} $ \\
     \hline
 $\phantom{ } \theta_{\rm E,bd}$     $[\arcsec]$ &  $ \phantom{-} 0.13$   & $  \phantom{-}0.13_{-0.02}^{+0.02}$ & $  \phantom{-} - $ & $  \phantom{-} - $ & $  \phantom{-} - $ & $  \phantom{-} 0.13_{-0.07}^{+0.08} $ & $  \phantom{-}0.25_{-0.04}^{+0.03} $  & $  \phantom{-} - $ \\  
 $\phantom{ } r_{\rm t,bd}$     $[\arcsec]$ &  $ \phantom{-} 33.50$  & $  \phantom{-}33.50_{-4.7}^{+1.5}$ & $  \phantom{-} - $ & $  \phantom{-} - $ & $  \phantom{-} - $ & $  \phantom{-} 23.00_{-3.20}^{+3.20} $ & $  \phantom{-}34.90_{-5.80}^{+3.40} $  & $  \phantom{-} - $ \\  
  \hline
 $\phantom{ } \rm rms$     $[\arcsec]$ &  $ \phantom{-} $   & $  \phantom{-}0.08 $ & $  \phantom{-}0.19 $ & $  \phantom{-}0.32 $ & $  \phantom{-}0.11 $ & $  \phantom{-}0.34 $ & $  \phantom{-}0.20 $ & $  \phantom{-} 0.45$  \\
\hline                                   
\end{tabular}}
\end{table*}

\subsubsection{MACS MP-full model results}
\label{sec:macs_mp}
In the case of the multi-plane modelling, where we model the parameters in the same set-up as the input, we recover, within the $ 1 \sigma$ uncertainties, the input parameters, as shown in Table \ref{tab:MACS_param}. The rms of this model is $0.08''$, which is very close to the observational uncertainty. We see that the offset between predicted and observed image positions is very small (Figure \ref{fig:MACSposoff}) and the distribution of predicted and observed magnification ratio is centred around 1 with very small scatter in the tails, as shown in Figure \ref{fig:MACSmagrat}. One of the images with high magnification (and therefore elliptical errors) is predicted to have a different image parity than observed. This is due to the fact that it is close to a critical curve, and a slight change on the model can change its relative position with respect to the curve. \\

\subsubsection{MACS SP1 model results}
\label{sec:macs_sp1}
If we model the input mass distribution keeping only the brightest foreground at the cluster redshift \citep[similarly to what was done in][]{Grillo2015, Caminha2016}, we find that the mass of this perturber is decreased. If we look at the scaled Einstein radii of the perturbers, as shown in Equation \ref{eq:scaled_thetaE}, using $z_{\rm s}=0.3$ (roughly the mean redshift), we see that the difference is less significant, but still the scaled Einstein radius at the real redshift is higher than that at the cluster redshift. 
This mass is compensated by an increase in the mass of the Southern halo and of the peculiar member (Figure \ref{fig:0416pecmember}) that has a freely varying mass profile (without scaling relation imposed) due to its proximity to the bright foreground, which is anyway not substantial. However, most of the parameters are recovered. The total rms is $\sim0.2''$, as shown in Table \ref{tab:MACS_param}. This model is however pretty good at predicting the image positions and magnification (magnification ratio median $\sim 0.97$ as shown in Table \ref{tab:MACSmagrat}), and does not show any parity flip in any of the predicted images.\\

\subsubsection{MACS SP model results}
\label{sec:macs_sp}

In the single plane model, which includes only the cluster, the lack of the LOS galaxies' mass is compensated by an increase in the mass of the cluster members and halos. This model also shifts the centroid position of the southern halo (becoming closer to the bright massive foreground perturber) and of the smaller halo by almost $1\arcsec$. 
The total rms of this model is $\sim0\farcs32$, which is very close to the model precision reached by the single-lens-plane models in recent years. Therefore, residual rms between the observed image positions and the image positions predicted by single-lens-plane models can actually be due to the lack of appropriate treatment of the lens environment \citep[as previously suggested by e.g., ][]{Jullo2010, DaloisioNatarajan2010}. Moreover, this shows that including the LOS galaxies at the wrong redshift performs better than not including them at all (Table \ref{tab:MACS_param} and Section \ref{sec:macs_sp1}). 
However, the reconstruction of magnification, and therefore the intrinsic brightness of the source, is still possible with the single plane models with an error of $\sim10\%$. Some more care is needed with images close to critical curves and highly magnified, some of which are predicted to have an opposite image parity in this model as well.\\

\subsubsection{MACS MP-fore model results}
\label{sec:macs_mpf}
If we include only the foreground objects at their correct redshift, we find a slight increase of mass in both the members and the halos, to compensate for the lack of the background perturbers, but we can still recover the input within $1-2 \sigma$, as shown in Table \ref{tab:MACS_param}. The rms of this model is $0\farcs11$, which is also quite close to the observational uncertainty. As in the MP-full model, we find that the offset between predicted and observed image positions is very small and the distribution of predicted and observed magnification ratio is centred around 1 with very small scatter in the tails, as shown in Figure \ref{fig:MACSmagrat} and Table \ref{tab:MACSmagrat}.  Moreover, the same image as in Section \ref{sec:macs_mp} is predicted with opposite image parity.\\

\subsubsection{MACS MP-back model results}
\label{sec:macs_mpb}
In the case of the multi-plane modelling with only the background galaxies, we find that the rms is $\sim0\farcs3$, as in the single plane model. We see that the centroid of the two massive halos are shifted by $\sim1\arcsec$, because the model did not take into account the lensing effect of the foreground galaxies. Moreover, to account for the lack of foreground, we find that the members and the southern halo are more massive. We note that this is probably due to the fact that the two most massive foreground galaxies are in the southern region of the cluster.  Magnification is reconstructed with $\sim10\%$ error, and this model predicts an image with flipped parity, which is very close to the Southern BCG, and thus more sensitive to the increase in mass of members and of the Southern halo. \\

\subsubsection{MACS MP-constML model results}
\label{sec:macs_mp_mlc}
\begin{figure*}[htbp]
   \centering
   \includegraphics[width=0.9\textwidth]{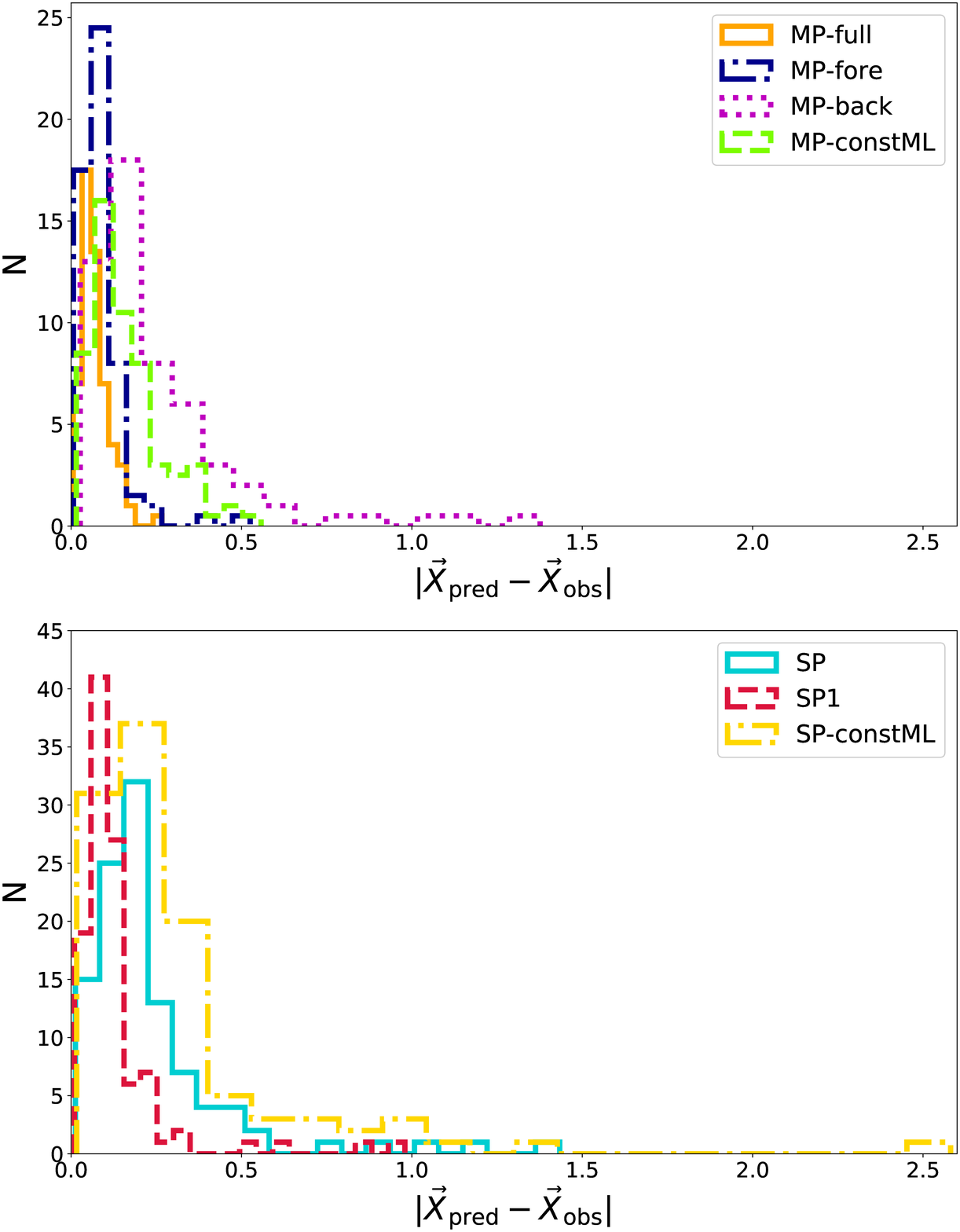} 
   \caption{Positional offset of the simulated observed vs predicted image positions of the different models of Mock MACS. The top panel shows the multi-plane models, and the bottom panel shows the single plane models. As expected, the full multi-plane model (black) is the one that reproduces the image position more closely. Accounting for the perturbers' mass, even if at the wrong redshift (model SP1, in red, and MPfore in green) reproduces the images better among the single plane models. The standard single-plane model with the wrong mass-to-light ratio (yellow) is the one with greater offsets.  }
   \label{fig:MACSposoff}
\end{figure*}

\begin{figure*}[htbp]
   \centering
   \includegraphics[width=0.9\textwidth]{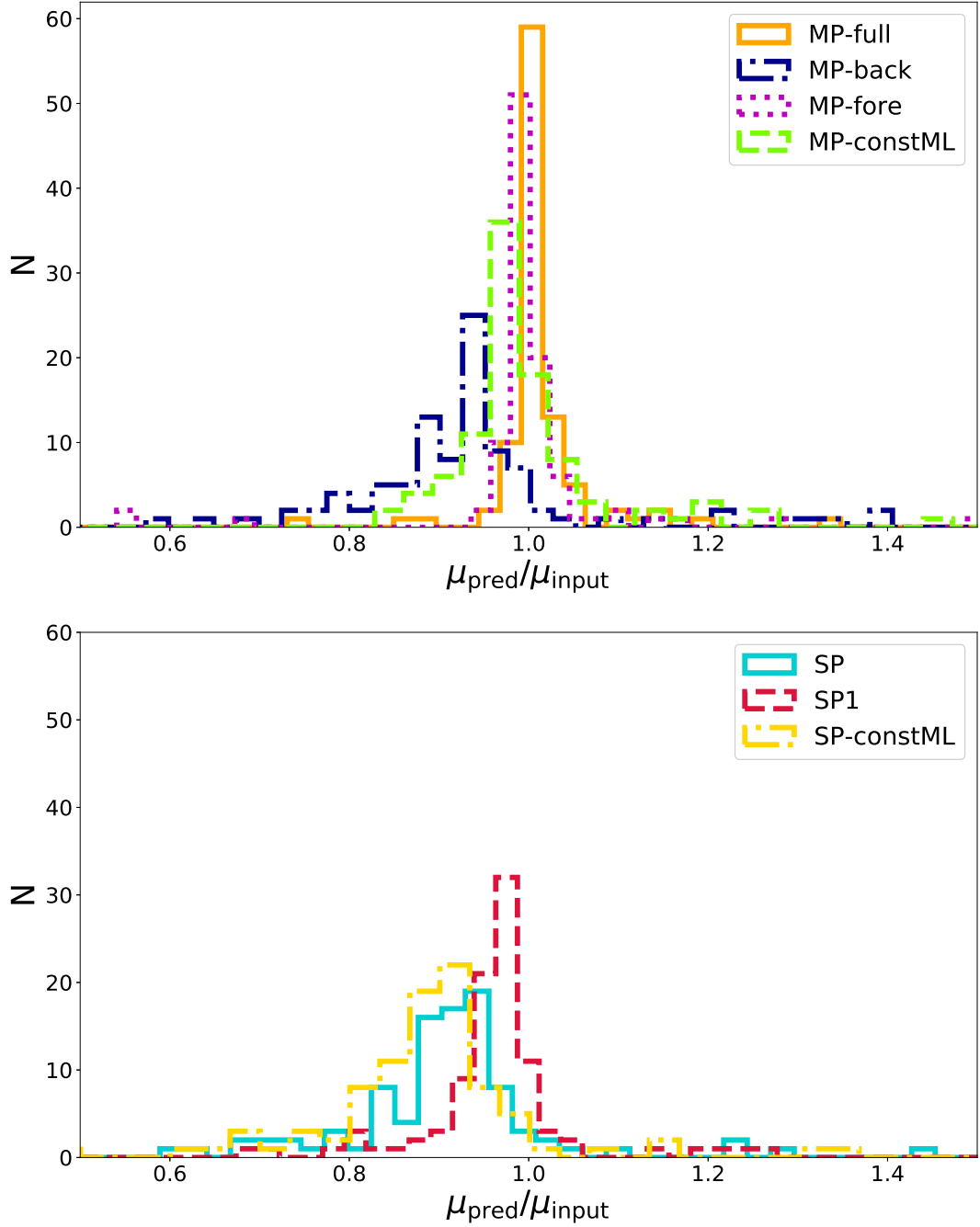} 
   \caption{Magnification ratio of the different Mock MACS models and the input Mock MACS. The top panel shows the multi-plane models, and the bottom panel shows the single plane models. Interestingly, some images are predicted with high magnification ratio, and some are predicted with a different parity. These images are all the arc-shaped images that lie close to the critical curves. This is due to the fact that model changes modify the shape of the critical curves, and therefore images that are close to those may change their relative positions with respect to them. }
   \label{fig:MACSmagrat}
\end{figure*}

If we scale the cluster members and LOS galaxies assuming a different mass-to-light ratio, as for example that of Equation \ref{eq:ML_constscale}, we find, as expected, that the offset in the predicted and observed image positions is greater than that of the multi-plane model, i.e. $0\farcs2$. As shown in Figure \ref{fig:MACSposoff} and \ref{fig:MACSmagrat}, the image positions and magnification are, overall, quite well reconstructed. However, in this case we have 5 images that show a flip in parity, one of which is very close to the Southern BCG and is very highly magnified, and the others are part of the system in the North-West region of the cluster, which shows 8 images of two clumps of the same source galaxy around two cluster members \citep[system 14 in ][ more detailed discussion in Section \ref{sec:MACSresultsmagn}]{Caminha2016}. \\

\subsubsection{MACS SP-constML model results}
\label{sec:macs_sp_mlc}
\begin{figure*}[htbp]
   \centering
   \includegraphics[width=\textwidth]{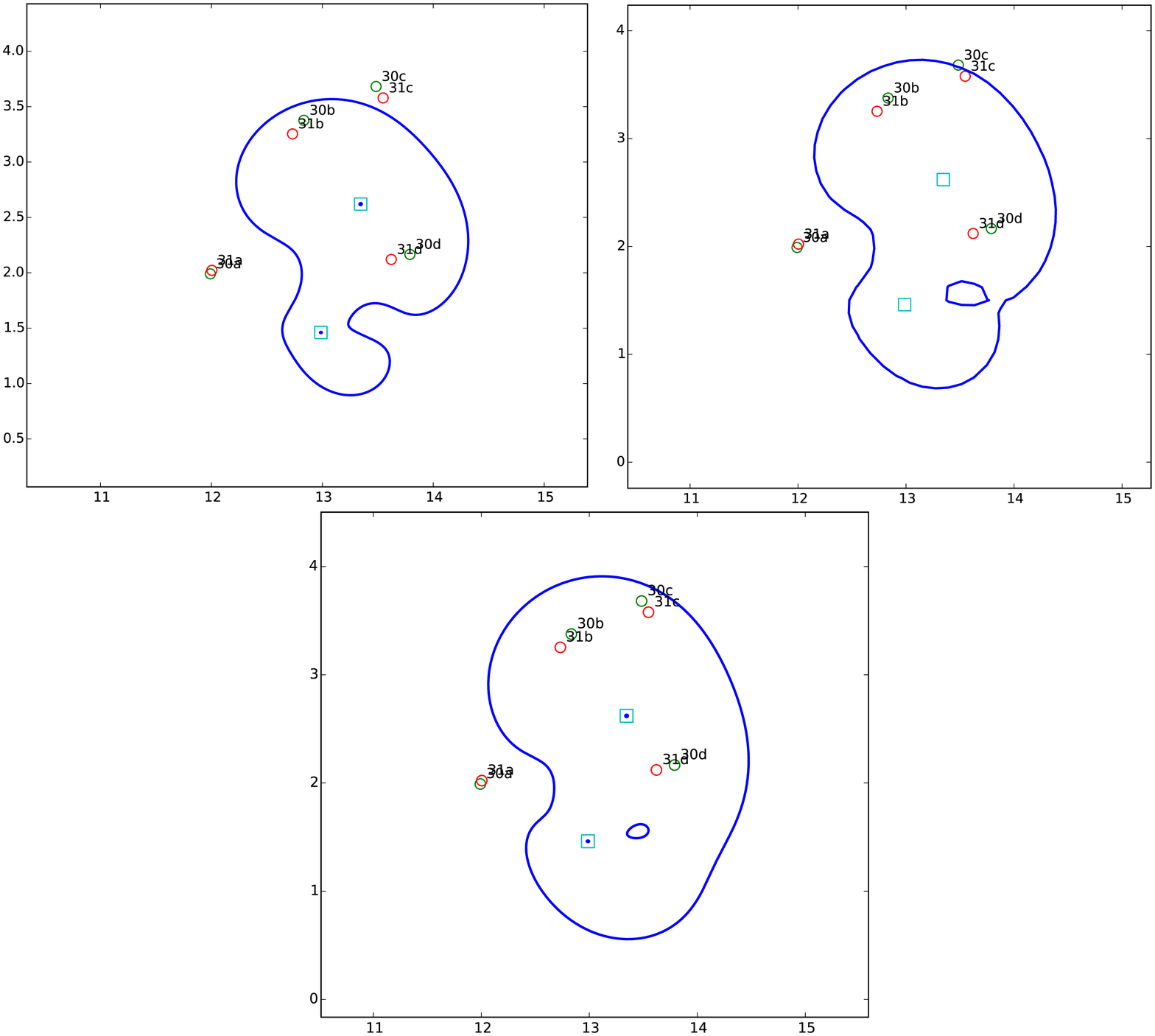} 
   \caption{Critical curves for the region of the multiple images of the source galaxy at $z_{\rm s}=3.2213$ of our Mock MACS models: MP-full (top-left), the MP-constML (top-right) and the SP-constML (bottom). The different clumps of the source are labelled with different numbers (30 and 31) We see that a change in the modelling assumptions can change quite significantly the shape of the critical curves, especially for images close to them.}
   \label{fig:MACS_3.2213}
\end{figure*}

If we model the input mass distribution with the single plane model and a constant mass-to-light ratio,  as in Equation \ref{eq:ML_constscale}, we find that the image positions are predicted with an overall rms of $\sim0\farcs45$. This model is, expectedly, the worst at reproducing the image positions and magnification, and even has 2 predicted pairs of images on the same side of the critical curves. These images belong to the system of multiple images already mentioned in Section \ref{sec:macs_mp_mlc} \citep[system 14 in][]{Caminha2016} that correspond to the same source at $\rm z=3.2213$. We show the critical curves of that region in Figure \ref{fig:MACS_3.2213} (bottom panel). and we discuss in Section \ref{sec:MACSresults}.
It also predicts two of the images with high magnification with a parity flip with respect to the input magnification. \\

\begin{table}[ht]
\caption{Magnification ratios between the reconstructed and input magnifications for different Mock MACS models. The listed values are the medians of the probability distribution functions of the magnification ratio, with $1\sigma$ uncertainties.}             
\label{tab:MACSmagrat}      
\centering                          
\renewcommand{\arraystretch}{1.4}  
\resizebox{\width}{!}{
\begin{tabular}{lcc}        
\hline                 
Model & $\mu_{\rm model}/ \mu_{\rm input} $  \\    
\hline                        
 MP-full &         $  1.00_{-0.01}^{+0.05}$ \\
 MP-fore &   $  1.00_{-0.02}^{+0.03}$\\  
 MP-back &  $  0.93_{-0.11}^{+0.07}$ \\  
 MP-constML &  $  0.98_{-0.05}^{+0.07}$ \\  
 SP &   $  0.92_{-0.10}^{+0.06}$\\  
 SP1 &   $  0.97_{-0.06}^{+0.04}$\\  
 SP-constML &  $  0.90_{-0.16}^{+0.07}$ \\  

\hline                                   
\end{tabular}}
\end{table}

\begin{figure*}[htbp]
   \centering
   \includegraphics[width=\textwidth]{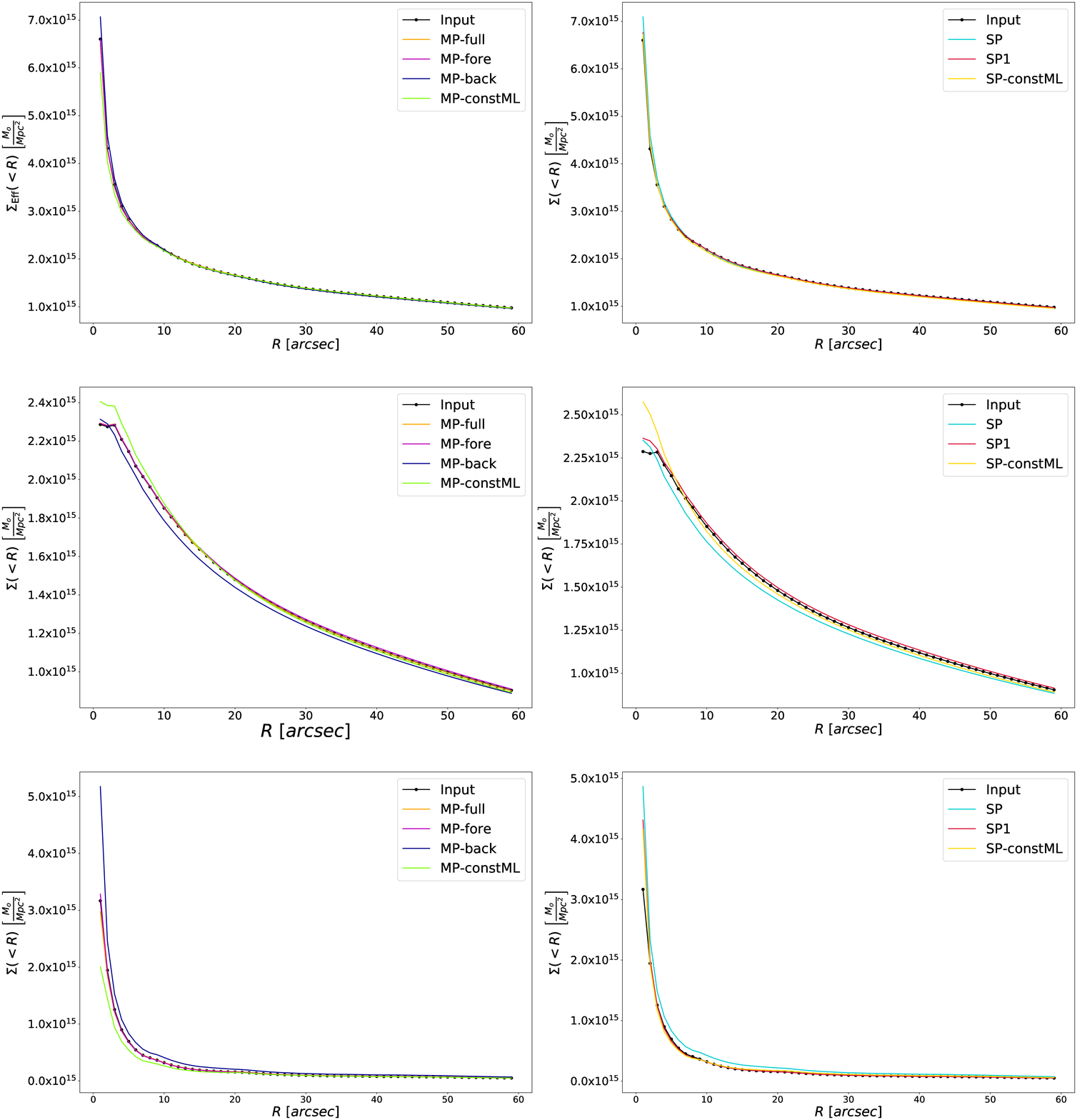} 
   \caption{Average surface mass density $\Sigma(<R)$ (for source at redshift $z_{\rm s}=3$)  as a function of radius (from G1) for the multi-plane MP (left panels) and single-plane SP (right panels) best-fit models of the Mock MACS mass distribution compared to the input model (in dotted black) . The upper panels show the $\Sigma_{\rm eff}(<R)$ for the total cluster, i.e. halos, members and perturbers, the central panels show the average surface mass density of the halos for the different models, and the bottom panels show the average surface mass density of the cluster members only. Note that in the total multi-plane configuration the $\Sigma_{\rm eff}(<R)$ is relative to the total deflection angle, as explained in Section \ref{sec:Mock2ASMD}.}
   \label{fig:MACSasmd}
\end{figure*}

\begin{figure*}[htbp]
   \centering
   \includegraphics[width=\textwidth]{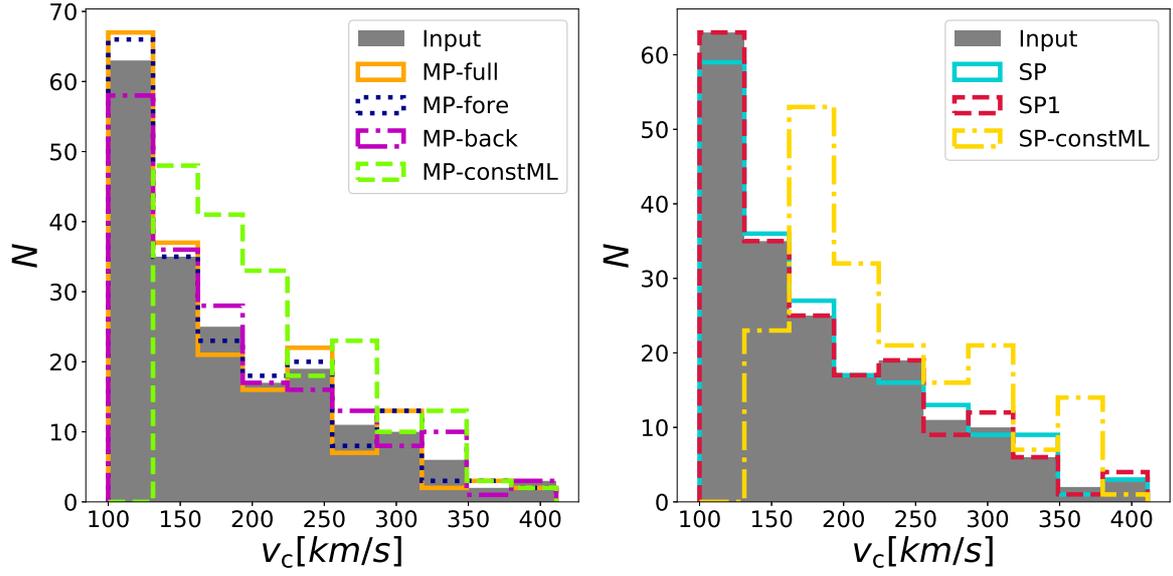} 
   \caption{Number of subhalos as a function circular velocity for our multi-plane (left) and single plane (right) models within an aperture of $\sim400\ \rm kpc$. Surprisingly, the model that reproduces better the input is the SP1 model, instead of the MP-full. However, we suspect this might be a coincidence specific to the configuration of \ourlens.}
   \label{fig:MACS_Nsigma}
\end{figure*}

\subsubsection{Results}
\label{sec:MACSresults}
As expected, the multi-plane model is the one that reproduces the image positions best, with a rms of around the observational uncertainty $\sim0\farcs 08$. However, accounting for the perturbers' mass, even if at the wrong redshift (as in model SP1 in Figure \ref{fig:MACSposoff}), allows the images positions to be reproduced better among the single plane models, with a rms of $\sim0\farcs2$. The standard single-plane model (SP) has a rms of $\sim0\farcs 32$. We therefore suspect that part of the offset in single plane models of galaxy clusters might actually be due to the exclusion of the perturbers. We also explore the effect of taking into account only the foreground and background galaxies, and we find that the inclusion of the foreground is more important for a better fit of the observables, confirming what was found for Mock mass distribution 1 (discussed in the Appendix) and in \citet{McCully2017}, for a more simplistic mock cluster. Actually, for the Mock MACS we find that including only the background gives a comparable rms to the single plane models, i.e. to not including any LOS galaxy at all. Modelling with the wrong scaling relation, as expected, increases the offsets of the predicted and observed image positions even more, for both the multi-plane and single plane model.

\subsubsection{Source magnification reconstruction}
\label{sec:MACSresultsmagn}
As for the image positions, the magnification is reconstructed with different precision by the different models. Figure \ref{fig:MACSmagrat} shows the magnification ratio (predicted vs. input) distribution for the different models. We find that the MP-full model distribution is centered around 1, showing that this model is the best at reproducing the input magnification. Also the MP-fore and SP1 models are able to reproduce the magnification quite well, with a median ratio of $\sim0.97-1$, as shown in Table \ref{tab:MACSmagrat}. All the other models tend to predict a lower magnification than the input. In Table \ref{tab:MACSmagrat} we list the median values of the magnification ratio distributions and the respective $1\sigma$ uncertainties (16th and 84th percentiles). However, all the distributions have median values within $0.90-1.00$, therefore the median error on the magnification reconstruction is still within $\sim10\%$. Interestingly, some image systems with high magnification and on (or close to) a critical curve, in almost all the models are predicted with a high magnification ratio and/or with a parity flip (we do not show these outliers in Figure \ref{fig:MACSmagrat} for visualization convenience). An example of system with high magnification ratio and parity flip is Figure \ref{fig:MACS_3.2213}. In this Figure we show the 8 simulated observed images of two source clumps of a galaxy at redshift $z_{\rm s}=3.2213$, which are located in the North-West region of the cluster \citep[system 14 in ][]{Caminha2016}. The multiple images of this source galaxy are located in the neighbourhood of two cluster members (cyan squares in Figure \ref{fig:MACS_3.2213}), that act as strong lenses for that particular source. The images are highly magnified and form arcs around those members, and they are very sensitive to the shape of the critical curves due to the presence of those members. If we look at the top-left panel of Figure \ref{fig:MACS_3.2213} (corresponding to the input model), we see that a pair of images (30c,31c) are predicted to be on different side of critical curves, with respect to the bottom panel (corresponding to the SP-constML model). Therefore, in the SP-constML model their predicted image positions coincide with those of, respectively, image 30b and 31b. If we compare the input with the top-right panel (MP-constML model), we see that in this case the images 30c and 31c are closer to the critical curves, so they are predicted with a much higher magnification. Indeed, as already shown in Section \ref{sec:Mock2constML}, changing the mass-to-light ratio with which we scale the Einstein and truncation radii of the cluster members, increases or decreases the mass of the members, and consequently changes the shapes of the critical curves for those members, affecting the relative position with respect to critical curves of images nearby. We point out that, in the model of \citet{Caminha2016}, these two cluster members were considered as free parameters instead of being scaled with the other members, since they are the main contributors to the creation of the multiple images of this system \citep[system 14 in ][]{Caminha2016}. In general, we suspect images close or on the critical curves need to be treated carefully when trying to reconstruct the intrinsic brightness of the background sources. 
 
 \subsubsection{Mass reconstruction}
\label{sec:MACS_massrec}
Figure \ref{fig:MACSasmd} shows the average surface mass density for our models. We do not observe significant differences among the different models for the total $\Sigma(<R)$, probably since the number of the perturbers is much smaller than the total cluster mass, therefore they do not contribute significantly to the total mass load. Moreover, all the models agree very well in the region of the Einstein radius of the cluster, since the mass within the Einstein radius is the quantity that lensing constraints tightly. 
If we look at the single matter components, i.e. only the halos (central panel) and only the members (bottom panel), we find slight differences among the models, especially in the inner part of the cluster, while in the outskirts they all agree very well. The models that are slightly different from the input are, among the MP models, the MP-back and MP-constML, in both the halos and cluster members $\Sigma(<R)$. The model MP-back, as we already saw from the positional offset (Table \ref{tab:MACS_param} and Figure \ref{fig:MACSposoff}), is the worst at reproducing the observables among the MP models. This tells us that including only the background perturbers can even perform similarly to the SP model, i.e. not including them at all. This implies that the average surface mass density can be similar to that of the SP model (as shown in Figure \ref{fig:MACSasmd}). Assuming a different mass-to-light relation makes the halos $\Sigma(<R)$ steeper within the inner part of the cluster, and the cluster members' one less peaky. We see the same trend in the SP case. Despite these slight differences, the overabundance of galaxies at the center of \macs\ in observations compared to simulations noted in \citet{Grillo2015} is robust against the presence of LOS perturbers.
Figure \ref{fig:MACS_Nsigma} shows the number of cluster members as a function of their circular velocity. If we compare it to the input (grey filled line), we see that most of the multi-plane and single plane models do not change substantially the overall velocity dispersion distribution of the members, showing that this is robust even though the model is not complete. However, we see that both in the single and multi-plane case, if we assume a wrong scaling relation, the distributions tend to prefer more massive galaxies (around 100 km/s in the multi-plane case and even 200 km/s in the single plane case). Thus, the choice of scaling relation could potentially alleviate the tension in the disparate numbers of massive galaxies and subhalos in the inner parts.  Encouragingly, our simulations suggest that the $\chi^2$ of the model fit could possibly be used to probe the underlying scaling relation.  It is worth exploring further scaling relations of the cluster galaxies in future models of \ourlens. \\

\section{Summary}
\label{sec:summary}
In this work we explored the effects of the LOS galaxies in strong gravitational lensing modelling of galaxy clusters. We simulated different galaxy clusters and their environment, building models of increasing complexity and realism, that we used to simulate strong lensing observables. We then determined the lensing halos' and galaxies' parameters with different assumptions and compared to the input simulated cluster to assess the effects of the LOS perturbers. \\
The simulated system Mock cluster mass distribution 2 is composed by a cluster at $\zc=0.4$ with a halo and ten elliptical galaxies having different, realistic luminosities, axis ratios and orientations. We assumed a total mass-to-light ratio corresponding to the tilt of the fundamental plane. We added two foreground perturbers at $z_{\rm fd}=0.2$, one close and one far away in projection, and one close-in-projection background perturber at $z_{\rm bd}=0.6$. All the perturbers have equal, large mass, random ellipticity (between $0.6$ and $1$) and orientation. We use this configuration to simulate mock lensing data, and we obtain a set of 17 multiple image positions of 3 background sources.
In this mock system we explored the effect of different mass-to-light relations and the spherical-elliptical galaxies assumption. We found that:
\begin{itemize}
\item[1.] Far-in-projection ($\sim100''$) perturbers do not affect substantially the other parameters' values. Indeed their posterior probability is sampled as flat and they do not look correlated to other parameters.
\item[2.] In multi-plane modelling, assuming spherical galaxies (both members and perturbers) recovers different profile parameters than the input, and has generically lower magnification. Despite this, the multi-plane reconstruction has a rms very close to the observational uncertainty. Therefore, one needs to be careful in interpreting the reconstructed mass distribution and magnification, as the goodness of fit does not allow to distinguish between spherical and elliptical galaxies. However, the difference in the parameters is less prominent when only the cluster members are treated as spherical. This might be due to the fact that the perturbers of this model are very massive and very close to the cluster center, therefore they have a large effect on the total mass reconstruction. Moreover, we suspect this effect might depend on the simplicity of our model and might be mitigated in more realistic clusters with higher number of cluster members.
\item[3.] Multi-plane models have a peakier average surface mass density than single plane models, which instead show a shallower profile. However, all the average surface mass density values match at $\rm R\sim10''$, which is the total Einstein radius of the cluster. This shows that the projected total mass enclosed within the Einstein radius, that is the quantity that strong gravitational lensing constraints tightly, is well reconstructed by all these models.

\end{itemize}
 
The other simulated cluster we studied is a realistic model of the HFF cluster \ourlens , labelled Mock MACS, which we built using 107 spectroscopically confirmed multiple images from 37 sources (with $1<z_{\rm s}<6$) as constraints, and then modelling the three halos (two located in the NE, and one in the SW direction), the 193 spectroscopically confirmed cluster members, and the LOS galaxies' profile parameters. We assume the cluster members and LOS galaxies to be spherical.   
We find that our mock of \ourlens\ which includes 11 LOS perturber galaxies is able to reproduce the real cluster's observables with a rms of $\sim0\farcs53$.
We then model this simulated Mock MACS cluster to assess the impact of LOS galaxies. Our results can be summarised as follows:
\begin{itemize}
\item[1.] Neglecting the cluster's LOS environment leads to a rms of $\sim0\farcs3$ in the offset distance between observed and modelled image positions.
\item[2.] The inclusion of LOS galaxies in the cluster modelling at the wrong redshift (i.e. at cluster redshift) reproduces the observed image positions with a rms of $\sim0\farcs20$.
\item[3.] We find that foreground perturbers have a more significant effect than the background. However, since in this case the discrepancy is more prominent (including only the background is comparable to not including it at all), we ascribe it also to the particular configuration of the cluster. 
\item[4.] Neglecting the lensing environment or assuming it at the wrong redshift does not affect the reconstruction of the magnification of background sources substantially (errors within $\sim10\%$). However, apart from the full multi-plane reconstruction, all the other models tend to underpredict the magnification. We also find that systems with high magnification, or in general close to critical curves, are more sensitive to the different assumptions on the modelling. Indeed, since strong lensing is a highly non linear effect in these regions, small changes in the parameter values can cause changes on the critical curves positions. These can lead to a much higher predicted magnification and may also cause a flip in image parity.
\item[5.] We do not observe a substantial difference in the average surface mass density of the cluster among the various mass models we have considered, probably because these perturbers are very small compared to the total cluster mass. Thus, the overabundance of galaxies at the center of \macs\ in observations compared to simulations noted in \citet{Grillo2015} seems robust against the presence of LOS perturbers. However, the mass function of the cluster members depends on the assumed scaling relations. The dependence can be partly mitigated by using the goodness of fit ($\chi^2$) of the multiple image positions, since the “true” scaling relations generally lead to lower $\chi^2$.
\item[6.] We find a correlation between the masses of the foreground line-of-sight perturber galaxies and the halos' centroid coordinates. We also see that the Einstein radius of the different mass components are anti-correlated among each other, since a decrease of mass in one can be compensated by an increase in another component. Moreover, the Einstein radius and truncation radius of the galaxies are correlated with each other.
\item[7.] Assuming different scaling relations can lead to very different results for the mass of the members and of the halos, and therefore can change the substructure distribution quite significantly. 

\end{itemize}

Finally, from our best-fit model of \ourlens\ we find that, for this particular cluster, the total $rms$ does not decrease substantially after the inclusion of LOS galaxies with respect to single plane models. This suggests that the dynamic and turbulent nature of this cluster is still probably not well depicted by e.g. the simplistic model of the halos. For future model improvements of this cluster those effects would need to be accounted for. We also find that using the correct scaling relations is crucial, especially for reconstructing the subhalo mass function. Therefore, we conclude that the main contribution to the residual $rms$ for this particular cluster is not due to LOS galaxy, and future models will have to go towards including more complex halo shapes, accounting for the ellipticity of galaxies and improving the scaling relations for the cluster members.

\begin{acknowledgements}
GC and SHS thank the Max Planck Society for support through the Max Planck Research Group for SHS. GC acknowledges Prof. Alexandre Refregier and Dr. Adam Amara for the helpful discussions. CG acknowledges support by VILLUM FONDEN Young Investigator Programme grant 10123. GBC, AM and PR acknowledge financial support from PRIN-INAF 2014 1.05.01.94.02.

\end{acknowledgements}

\bibliographystyle{aa}
\bibliography{ms}

\section*{Appendix}
 
\subsection*{Mock Cluster lensing mass distribution 1}
\label{sec:Mock1}

\subsubsection*{Input}
\label{sec:Mock1input}
\begin{figure}
  \centering
   \includegraphics[width=1.0\columnwidth]{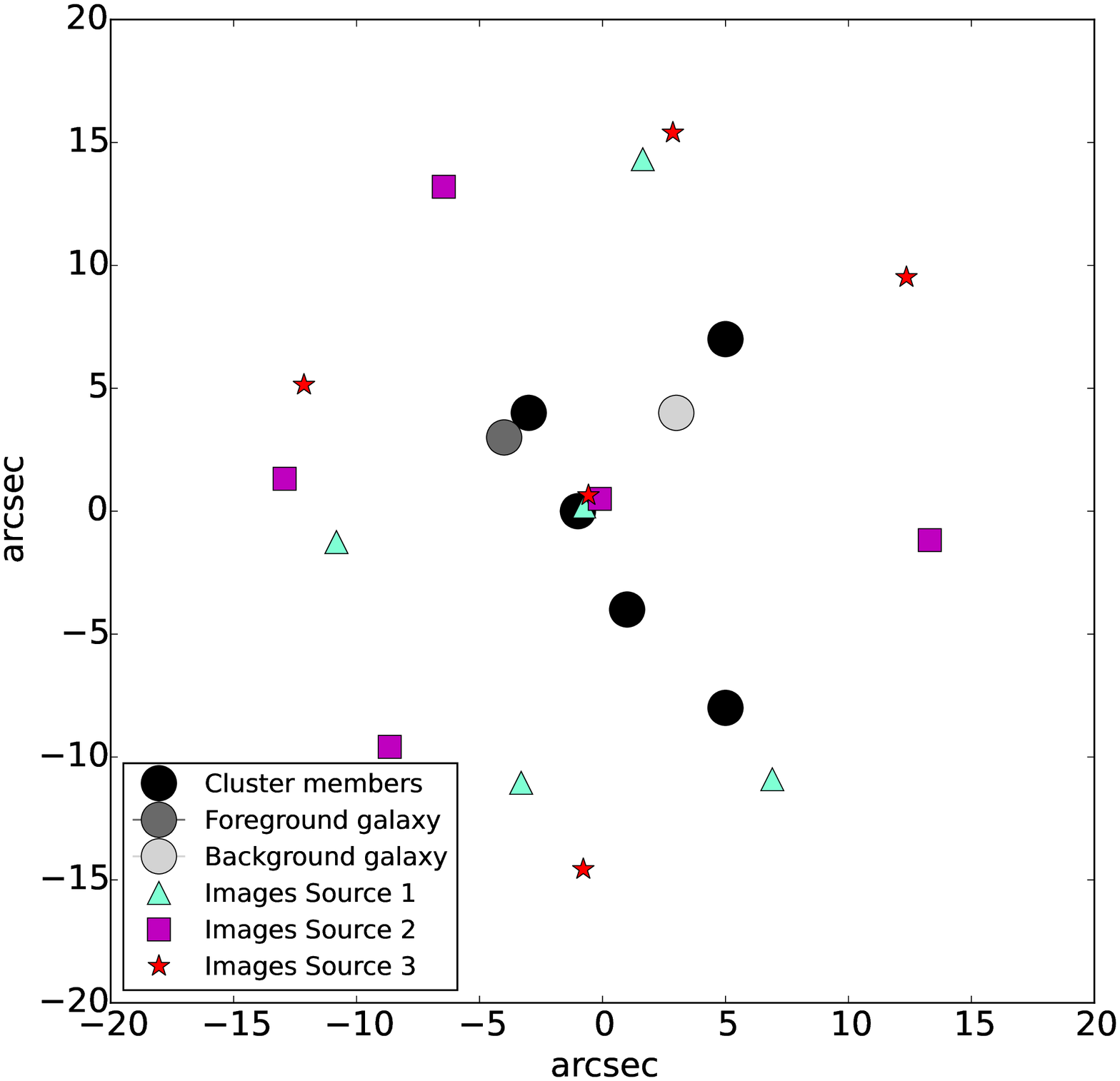} 
   \caption{Mock cluster lensing mass distribution 1. The black circles represent the lenses (cluster members), the grey circles the foreground galaxy (darker grey) and the background galaxy (lighter grey). The cyan triangles, magenta squares and red stars represent the images of the three sources, respectively at $z_{\rm s1}=1.5$, $z_{\rm s2}=2.0$ and $z_{\rm s3}=2.5$.}
   \label{fig:mock1_setup}
\end{figure}

The simulated mock cluster lensing mass distribution 1 is that of a cluster at redshift $\zc=0.4$ and two galaxies, a foreground galaxy at $z_{\rm fd}=0.2$ and a background galaxy at $z_{\rm bd}=0.6$, that are close in projection to the cluster and that we treat as perturbers. The cluster is composed of 5 galaxies and a dark matter halo. The input parameters are summarised in Table \ref{tab:Mock1_param}. We assume all the galaxies in the cluster to have the same luminosity, therefore same mass, for simplicity, and same truncation radius, that we fix to $15''$. We put 3 different point sources (S1, S2, S3) at different redshifts ($z_{\rm s1}=1.5,\ z_{\rm s2}=2,\ z_{\rm s3}=2.5$) and simulate their image positions in the case of the multi-plane configuration. We find that each source is mapped into 5 images, for a total of 15 images, and use these image positions as our observables. The set up of this model is illustrated in Figure \ref{fig:mock1_setup}.

\subsubsection*{Full multi-lens-plane modelling}
\label{sec:Mock1MP}
We optimise the free parameters of the multi-plane model (Einstein radius of the foreground, background and cluster galaxies, and all the parameters of the halo profile, i.e. Einstein radius, position, ellipticity, orientation, core radius and slope) by maximising the likelihood in the image plane. We use simulated annealing to find the global minimum and recover the best-fit parameter values. In this case, since we simulated the image positions with the multi-plane model, we recover, within the uncertainties estimated by MCMC sampling, the initial parameters we have used to simulate, as shown in Table \ref{tab:Mock1_param}. The modelled image positions and the magnifications are perfectly fitted, i.e. with a null total-rms offset. We find a strong correlation between the mass of the halo and its axis ratio and between the core radius of the halo and the slope of the halo profile. We also find a strong anti-correlation between the centroid position of the halo and the mass of the foreground galaxy, which is explained by the fact that the centroid position is lensed by the foreground galaxy, and between the mass of the cluster galaxies and the axis ratio of the halo. We also find that the mass of the background lens is strongly anti-correlated with the halo mass, core radius and slope, that could also be explained as a lensing effect by the cluster. 
\subsubsection*{Single-plane modelling}
\label{sec:Mock1SP}
We then remove the foreground and background lens and fit the same image positions with the cluster alone. We vary the cluster parameters and find, as shown in Table \ref{tab:Mock1_param}, that the halo is still oriented along on the $x$ direction, and the ellipticity is recovered within the errors. However, its centroid position is offset of $\sim1''$ in both $x$ and $y$ direction, its mass is bigger and its core radius is larger by $4''$, making the profile less peaky in the center. Moreover, we recover a smaller mass for the cluster galaxies. 
In this case we find an additional strong anti-correlation between the halo and the cluster galaxies Einstein radius, which was not so prominent in the multi-plane system. This could be explained as the removal of the two LOS perturbers influences the distribution of the mass between the two remaining mass components, i.e. the halo and the cluster galaxies. Moreover, we find degeneracies between the axis ratio and the Einstein radius of the cluster and between the latter and the cluster galaxies' Einstein radius, since these degeneracies would keep approximately the same total mass enclosed within the multiple images, that is what strong lensing constrains tightly.
As for the image positions, the single lens systems predicted images are all offset by $<0\farcs8$ in both x and y direction. In terms of total-rms, the image offset is $\sim0\farcs4$. The magnification of the single plane configuration appears to be in general different, within a ratio of 0.8 to 1.8 with respect to the input. 

\subsubsection*{Cluster and single perturber modelling}
\label{sec:Mock1singlepert}
To investigate the effects of the single perturbers we model the same system with the cluster and the foreground lens only and the cluster and the background lens only, respectively.
We find that the addition of the foreground perturber to the single-plane model allows us to recover parameters that are more similar to that of the multi-plane model. Indeed, as shown in Table \ref{tab:Mock1_param}, the halo centroid position is offset by only $\sim0\farcs25$, its mass is not as big as that of the single-plane model, and the halo slope is recovered within the error. If we include only the background perturber, instead, we get the same parameter values, within the errors, as that of the single plane model, confirming results from previous studies \citep[e.g][]{McCully2014}, which showed that foreground perturbers have a more significant effect on the modelling compared to the background ones.
In terms of image positions rms, we find a total rms of $0\farcs18$ for the multi-plane model with only the foreground galaxy, while a total rms of $\sim0\farcs4$ for the model with only the background galaxy, showing that the addition of the foreground perturber is more significant for a good fit of the image positions.

\subsubsection*{Effect of individual perturber}
\label{sec:Mock1indivpert}
Before moving to a more realistic model, we perform a test to further investigate the effect of the individual perturbers along the line-of-sight. We use the Mock cluster lensing mass distribution 1 and we increase the mass of the background perturber such that the scaled Einstein radii of the two perturbers, for the intermediate source at $z_{\rm s2}=2$, is equal, namely
\be
\label{eq:scaled_thetaE}
\scalebox{1.2}{$ \frac{D_{\rm fd-s2}}{D_{\rm s2}} \theta_E^{\rm fd}=\frac{D_{\rm{bd-s2}}}{D_{\rm s2}} \theta_E^{\rm bd}.$}
\ee
We resimulate the three sets of multiple image positions using equation (17). We then repeat the analysis done in Section \ref{sec:Mock1singlepert} and we find that the perturber in the foreground is still more significant (rms $\sim0\farcs32$) than the background (rms $\sim0\farcs38$), but this effect is less prominent. However, in order for the scaled deflection angle of foreground and background galaxy to match, the background mass should be substantially larger than the foreground, i.e. with an Einstein radius of $\sim1\farcs$ bigger than that of the foreground.  So for typical scenarios where foreground galaxies have similar (or higher) masses as background galaxies, we should pay more attention to the foreground lenses.
 
\subsubsection*{Truncation radius}
\label{sec:Mock1rtrunc}
We try to model the truncation radii of the galaxies, i.e. roughly the half-mass radius \citep{EliasdottirEtal07}. At first we allow the truncation radii of all the cluster galaxies (which we assume to be equal) and of the two perturber galaxies to vary. We find that they are actually very difficult to constrain, even if the number of free parameters is significantly fewer than that of the constraints. We then fix the truncation radii for the two perturber and model varying the truncation radius only for the cluster galaxies. In this case we find that the truncation radius posterior has a flat distribution. We therefore suspect that its value is not really affecting the parameter modelling, as long as it is not significantly different from its true value. To explore this further we model the multi-plane system keeping all the truncation radii fixed to the wrong values ($10''$ instead of the input value $15''$). This shows that the wrong choice of truncation radii makes the estimation of the halo mass wrong by $\sim4\sigma$ and that of the cluster galaxies by $\sim2\sigma$, so non-negligible, as shown in Table \ref{tab:Mock1_param}. 

\subsubsection*{Conclusions}
\label{sec:Mock1conclusions}
The simulated system Mock cluster mass distribution 1 is composed by a cluster, at $z_{\rm c}=0.4$, with 5 spherical galaxies of equal luminosity and mass, and one dark matter halo with an Einstein radius of $10''$. Its environment is constituted by two close-in-projections LOS galaxies, one foreground $z_{\rm fd}=0.2$ and one background $z_{\rm bd}=0.6$, with equal, large mass. Our observables are a set of 15 images from 3 sources at redshifts, respectively $z_{\rm S1}=1.5,\ z_{\rm S2}=2.0,\ z_{\rm S3}=2.5$. On this cluster we studied the different effects of foreground and background perturbers. We conclude that:
\begin{itemize}
\item[1.] Foreground perturbers have a more significant effect in the modelling than the background, for a given equal mass, and their inclusion reproduces the observed image positions more accurately. This might be due to their lensing effects on the observed image positions. Moreover, foreground perturbers have lensing effects on the halos, affecting their centroid position and shape. If we try to equate the scaled Einstein radius (Equation \ref{eq:scaled_thetaE}) of foreground and background perturbers we see that this trend is slightly attenuated, but still visible.
\item[2.] The truncation radius of the perturbers does not affect substantially the lensing model. Indeed its posterior distribution appears flat and does not look correlated to the parameters of other profiles.
\end{itemize}  
This is also confirmed by the Mock MACS model, as discussed is Section \ref{sec:MACSresults}.
\begin{table*}[ht]

\caption{Constraints on lens parameters for different models of Mock cluster lensing mass distribution 1. The first column refers to the values used to simulate, the other columns refer to different models, in order, the full multi-plane, the single cluster-plane, the multi-plane with only foreground perturber, the multi-plane with only background perturber and the total multi-plane with the truncation radii of the galaxies fixed to the wrong value ($10''$ instead of $15''$).  The values are the medians of the posterior probability distributions of the lens parameters together with their 1$\sigma$ uncertainties. The orientation is measured counter clockwise from positive x-axis.  }             
\label{tab:Mock1_param}      
\centering                          
\renewcommand{\arraystretch}{2.2}  
\resizebox{\width}{!}{
\begin{tabular}{lcccccc}        
\hline                 
Parameters & Input & MP-full   & SP    & MP-fore & MP-back & MP-wt \\    

\hline                        
\hline
 $\phantom{ }\theta_{\rm E,\ fd}$        $[\arcsec]$ & $  \phantom{-}2.00$ &  $ \phantom{-} 1.99_{- 0.07}^{+ 0.07}$ & $  \phantom{-}{ - }$& $  \phantom{-}1.60_{- 0.20}^{+ 0.19}$   &  $ \phantom{-} {-}$  &$  \phantom{-}2.30_{- 0.08}^{+ 0.08}$\\
  \hline 
 $\phantom{ } x_{\rm halo}$        $[\arcsec]$ & $   0.0$ &  $ \phantom{-}0.01_{-0.05}^{+0.05}$ &  $ \phantom{-}0.98_{-0.08}^{+0.08}$  &  $ \phantom{-}0.21_{-0.11}^{+0.11}$ &  $ \phantom{-}1.15_{-0.09}^{+0.09}$ &  $ \phantom{-}-0.04_{-0.05}^{+0.05}$ \\
 $\phantom{ } y_{\rm halo}$        $[\arcsec]$ & $   0.0$ &  $ 0.00_{-0.05}^{+0.05}$ &  $ \phantom{-} 1.12_{-0.06}^{+0.07}$  &  $ \phantom{-}0.25_{-0.12}^{+0.12}$ &  $ \phantom{-} 1.07_{-0.07}^{+0.08}$ &  $ \phantom{-} -0.04_{-0.05}^{+0.05}$ \\
 $\phantom{ } \frac{b}{a}_{\rm halo}$                 & $   \phantom{-}0.8$ &  $ \phantom{-}0.80_{-0.01}^{+0.01}$  &  $ \phantom{-}0.84_{-0.03}^{+0.02}$ &  $ \phantom{-}0.83_{-0.02}^{+0.01}$ &  $ \phantom{-} 0.85_{-0.02}^{+0.02}$ &  $ \phantom{-} 0.82_{-0.01}^{+0.01}$ \\
$\phantom{-}\theta_{\rm halo}$  $[rad]$   & $  0.0$  &  $ 0.00_{-0.00}^{+0.00}$ &  $ 3\pi_{-0.01}^{+0.01}$ &  $ \pi_{-0.01}^{+0.01}$ &  $ -0.97\pi_{-0.01}^{+0.01}$ &  $ -0.00_{-0.00}^{+0.00}$ \\
 $\phantom{ } \theta_{\rm E,\ halo}$     $[\arcsec]$ & $   \phantom{-} 10.00$ &  $ \phantom{-} 10.00_{-0.04}^{+0.04}$ &  $ \phantom{-} 14.10_{-0.77}^{+0.73}$ &  $ \phantom{-} 12.80_{-0.47}^{+0.46}$ &  $ \phantom{-} 14.10_{-0.89}^{+0.86}$ &  $ \phantom{-} 11.40_{-0.32}^{+0.31}$  \\
 $\phantom{ } r_{\rm c,\ halo}$        $[\arcsec]$ & $  \phantom{-} 2.00$ &  $\phantom{-} 2.00_{-0.35}^{+0.35}$ &  $\phantom{-}6.1_{-1.2}^{+0.93}$ &  $\phantom{-}3.40_{-0.99}^{+1.60}$ &  $\phantom{-}6.60_{-1.30}^{+1.10}$ &  $\phantom{-}2.40_{-0.37}^{+0.39}$ \\
 $\phantom{ }\gamma_{\rm halo}$         & $   0.4$ &  $ 0.40_{-0.02}^{+0.02}$ &  $ 0.70_{-0.09}^{+0.07}$ &  $ 0.49_{-0.06}^{+0.10}$ &  $ 0.69_{-0.11}^{+0.08}$ &  $ 0.41_{-0.02}^{+0.02}$ \\
 $\phantom{ }\theta_{\rm E,\ g}$      $[\arcsec]$           & $   \phantom{-}2.00$ &  $ \phantom{-}2.00_{-0.05}^{+0.05}$ &  $ \phantom{-}2.00_{-0.20}^{+0.24}$ &  $ \phantom{-}1.88_{-0.12}^{+0.14}$ &  $ \phantom{-}1.80_{-0.19}^{+0.22}$ &  $ \phantom{-}2.13_{-0.05}^{+0.05}$ \\
    \hline 
 $\phantom{ }\theta_{\rm E,\ bd}$     $[\arcsec]$ & $  \phantom{-}2.00$  & $  \phantom{-}2.00_{- 0.24}^{+ 0.23}$ &  $ \phantom{-} {-}$  &  $ \phantom{-} {-}$ & $  \phantom{-}1.50_{- 0.61}^{+ 0.55}$& $  \phantom{-}2.02_{- 0.20}^{+ 0.22}$  \\
     \hline
     \hline
 $\phantom{ } \rm rms$     $[\arcsec]$ &  $ \phantom{-} $ & $  \phantom{-} 8\times10^{-4}$  &  $ \phantom{-} 0.41$ & $  \phantom{-} 0.18$  & $ \phantom{-}0.39$ & $ \phantom{-}0.028$ \\  
\hline                                   
\end{tabular}}
\end{table*}

\end{document}